\documentclass[aps,prb,twocolumn,superscriptaddress,letterpaper,floatfix]{revtex4-1}
\usepackage{graphicx}  
\usepackage{dcolumn}   
\usepackage{bm}        
\usepackage{amssymb}   
\usepackage{placeins}
\usepackage{dcolumn}


\begin{document}

\date{\today}
\title{Separation of magnetic and superconducting behaviour in YBCO6.33 (T$_c$=8.4 K)}

\author{Zahra Yamani}\email{zahra.yamani@nrc.gc.ca}
\affiliation{Canadian Neutron Beam
Centre, Chalk River Laboratories, Chalk River, ON K0J 1J0,
Canada}

\author{W.J.L. Buyers}
\affiliation{Canadian Neutron Beam
Centre, Chalk River Laboratories, Chalk River, ON K0J 1J0,
Canada} 

\author{F. Wang}
\affiliation{Department of Physics, University of Toronto, Toronto,
ON M5S 1A7, Canada }

\author{Y-J. Kim}
\affiliation{Department of Physics, University of Toronto, Toronto,
ON M5S 1A7, Canada }

\author{J.-H. Chung}
\affiliation{Department of Physics, Korea University, Seoul 136-713, South Korea}

\author{S. Chang}
\affiliation{NIST Center for Neutron Research, Gaithersburg,
Maryland 20899, USA}

\author{P.M. Gehring}
\affiliation{NIST Center for Neutron Research, Gaithersburg,
Maryland 20899, USA}

\author{G. Gasparovic}
\affiliation{NIST Center for Neutron Research, Gaithersburg,
Maryland 20899, USA}

\author{C. Stock}
\affiliation{School of Physics and Astronomy, University of Edinburgh, Edinburgh, EH9 3JZ, UK}

\author{C.L. Broholm}
\affiliation{Department of Physics and Astronomy, Johns Hopkins
University, Baltimore, MD 21218, USA}

\author{J.C. Baglo}
\affiliation{Department of Physics and Astronomy, University of
British Columbia, Vancouver, BC V6T 1Z1,
Canada}\

\author{Ruixing Liang}
\affiliation{Department of Physics and Astronomy, University of
British Columbia, Vancouver, BC V6T 1Z1,
Canada}\affiliation{Canadian Institute for Advanced Research,
Toronto, ON M5G 1Z8, Canada}

\author{D.A. Bonn}
\affiliation{Department of Physics and Astronomy, University of
British Columbia, Vancouver, BC V6T 1Z1,
Canada}\affiliation{Canadian Institute for Advanced Research,
Toronto, ON M5G 1Z8, Canada}

\author{W.N. Hardy}
\affiliation{Department of Physics and Astronomy, University of
British Columbia, Vancouver, BC V6T 1Z1,
Canada}\affiliation{Canadian Institute for Advanced Research,
Toronto, ON M5G 1Z8, Canada}

\begin{abstract}

Neutron scattering from high-quality YBa$_2$Cu$_3$O$_{6.33}$ (YBCO6.33) single crystals with a T$_c$ of 8.4 K shows no evidence of a coexistence of superconductivity with long-range antiferromagnetic order at this very low, near-critical doping of $p$$\sim$0.055. However, we find short-range three dimensional spin correlations that develop at temperatures much higher than T$_c$. Their intensity increases smoothly on cooling and shows no anomaly that might signify a N\'{e}el transition. The system remains subcritical with spins correlated over only one and a half unit cells normal to the planes. At low energies the short-range spin response is static on the microvolt scale. The excitations out of this ground state give rise to an overdamped spectrum with a relaxation rate of 3 meV. The transition to the superconducting state below T$_c$ has no effect on the spin correlations. The elastic interplanar spin response extends over a length that grows weakly but fails to diverge as doping is moved towards the superconducting critical point. Any antiferromagnetic critical point likely lies outside the superconducting dome. The observations suggest that conversion from N\'{e}el long-range order to a spin glass texture is a prerequisite to formation of paired superconducting charges. We show that while $p_c$ =0.052 is a critical doping for superconducting pairing, it is not for spin order.

\end{abstract}
\maketitle

\section{Introduction}
\label{labelOfFirstSection}

\begin{figure}
\begin{center}
\vskip 0cm
\resizebox{1\linewidth}{!}{\includegraphics{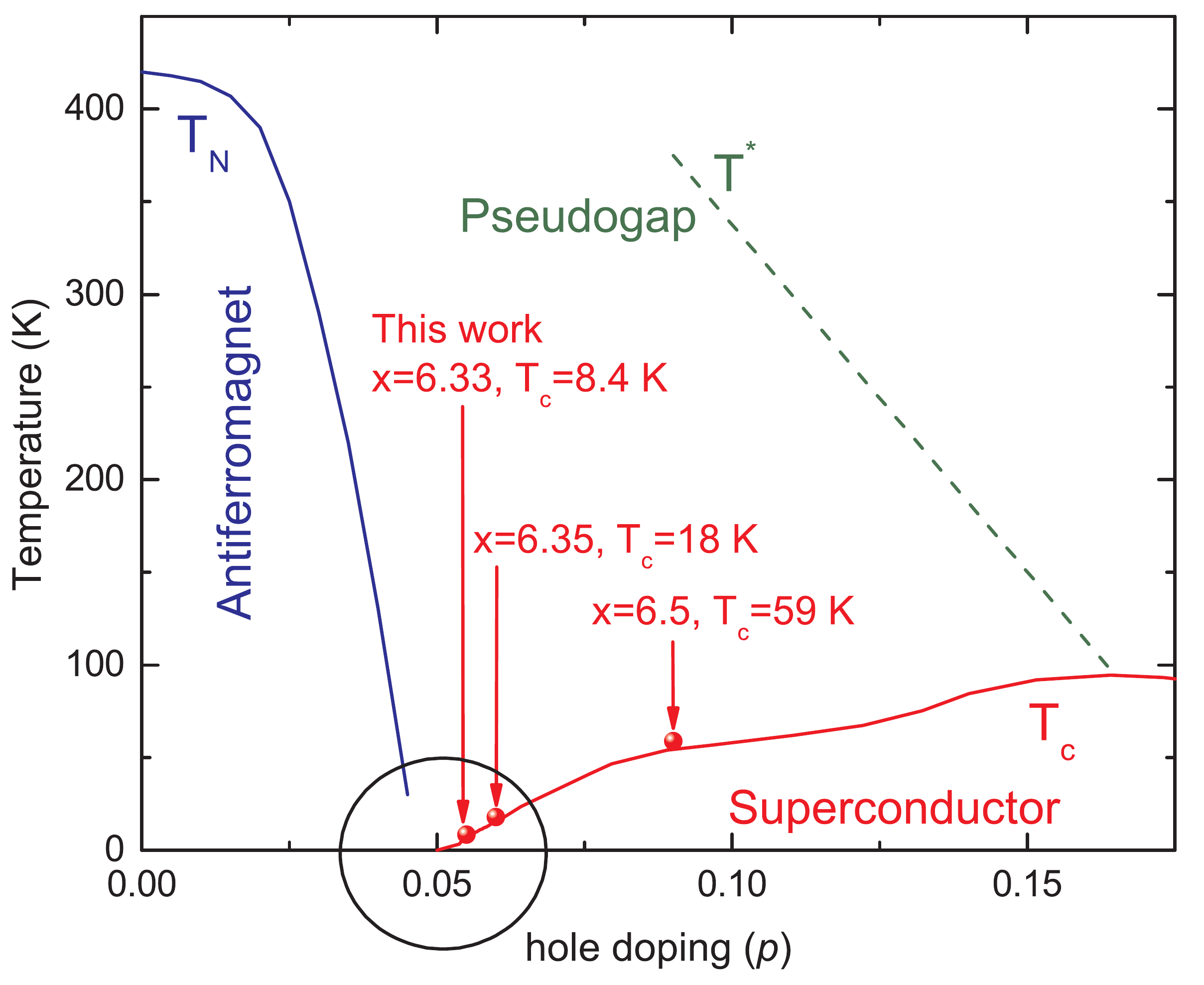}}\vskip
0cm \caption []{Electronic phase diagram of YBCO6+x is depicted as a function of hole doping. For the undoped parent material, there is an unambiguous transition to a long range 3D ordered AF phase at T$_N$. For $p > p_c$= 0.052 the system exhibits superconductivity below T$_c$. How the long range AF order gives way to superconductivity by doping remains a central question in the physics of HTSC cuprate supercoductors.   } \label{pdia}
\end{center}
\end{figure}

Despite numerous experimental and theoretical studies since the discovery~\cite{bednorz86} of high temperature superconductivity (HTSC) in cuprates in 1986, there is still no consensus on the superconducting mechanism in these materials. With enormous advances in growth of high-quality single crystals, however, a qualitatively common phase diagram has emerged (Fig.~\ref{pdia}). The phase diagram exhibits different states of matter, many of which are revealed by neutron scattering~\cite{fujita12,hayden08,tranq07,lee06,li08,phillips09,kastner98,bourges98,regnault98,rossat93,birg06,keimer08,keimer10,keimer10a,haug09,statt99,yamani07,stock04,li08,stock05,stock10,stock06,stock08,buyers06}. Only when novel properties of HTSC cuprates are fully elucidated across the whole phase diagram will a full understanding of these materials be gained. The search for new phases of matter is particularly important in the low doping region close to critical doping for superconductivity, $p_c$, because it creates the greatest challenge for theory.

The undoped HTSC parent materials are believed to be Mott-insulators with static long-range three dimensional (3D) antiferromagnetic (AF) order below a N\'{e}el temperature T$_N$. The Cu$^{2+}$ spins are localized with strong AF exchange coupling within 2D CuO$_2$ planes. The weak exchange coupling in the perpendicular direction allows the 3D N\'{e}el order to occur at a finite T$_N$. Initially, hole doping disrupts the network of correlated, localized Cu$^{2+}$ spins and suppresses antiferromagnetism. Further hole doping leads to superconductivity for doping larger than the critical doping $p_c$ $\sim$0.05 (underdoped region). The superconducting transition temperature T$_c$ increases with increasing doping and reaches a maximum at an optimal doping beyond which T$_c$ declines. Even though long-range AF order does not coexist with superconductivity in this region, strong spin fluctuations nonetheless survive well into the superconducting (SC) phase. Electronic properties more like conventional metals are exhibited at very high doping (overdoped region) but also in the underdoped region by the observation of quantum oscillations and Fermi surface reconstruction~\cite{vignolle12} in the range 0.08$<$$p$$<$0.15. Our results address a much lower doping region where superconductivity is weak and spin fluctuations are strong.

The single-layer La$_{2-x}$Sr$_x$CuO$_4$ (LSCO) cuprate family has been extensively studied by neutron scattering~\cite{kofu09,moho06,fujita02, matsuda02,waki01,matsuda00,matsuda00a,waki00,waki99,yamada98,keimer92}. The doped LSCO system intrinsically suffers from disorder very close to the CuO$_2$ planes in which both SC and AF occur. This is because doping involves substitution of La with Sr ions of different charge in layers directly adjacent to the CuO$_2$ planes. This chemical substitution inevitably introduces disorder into the CuO$_2$ planes which in turn distorts the CuO$_2$ planes. %
For the bilayer YBa$_2$Cu$_3$O$_{6+x}$ (YBCO6+x) system, where large and high-quality single crystals are available, doping occurs by varying the oxygen content in the CuO chains which are located far from the CuO$_2$ planes. The YBCO6+x system thus provides an intrinsically less disordered environment, where holes are introduced in the conducting plane by out-of-plane oxygen doping, that can reveal how doping affects the planar magnetic and superconducting properties.

Muon spin rotation ($\mu$SR) studies of underdoped YBCO6+x have led to the claim~\cite{sanna04, sanna10,niedermayer98} that a transition takes place to N\'{e}el order within the superconducting dome. As well, a transition to a frozen glass is claimed at an even lower temperature. One must emphasize that the $\mu$SR technique is a local probe whereas neutron scattering allows one to determine the length and time scales of the magnetic structure as it evolves with temperature and doping. Since there is little consensus on the nature of the precursor phase from which the SC phase emerges, we have undertaken a systematic neutron scattering study of high-quality YBCO6+x single crystals that lie close to the critical doping for superconductivity (see Fig.~\ref{pdia}). Our earlier experiments on YBCO6.35 (T$_c$=18 K, $p$=0.06) revealed that no long-range Bragg-ordered AF phase coexists with superconductivity~\cite{stock06,stock08,buyers06}. The short-range static spin correlations observed for YBCO6.35 leaves open the possibility that an AF quantum critical point or novel phase might occur at lower doping. We have therefore studied an even lower doped YBCO6+x, namely YBCO6.33 with T$_c$ of only 8.4 K and doping $p$=0.055 near the critical $p_c$=0.052. We will present clear evidence for the existence of spatially highly-correlated and slow spin fluctuations. Although the observations are qualitatively similar to higher doped YBCO6.35 (T$_c$=18 K)~\cite{stock06,stock08,buyers06}, these are the first measurements on YBCO6.33 with such a low T$_c$, and we will show that even for such low doped yet superconducting materials, long-range antiferromagnetic order and superconductivity do not coexist.

\section{Experiment}

\begin{figure}[tbh!]
\begin{center}
\vskip 0cm
\resizebox{1\linewidth}{!}{\includegraphics{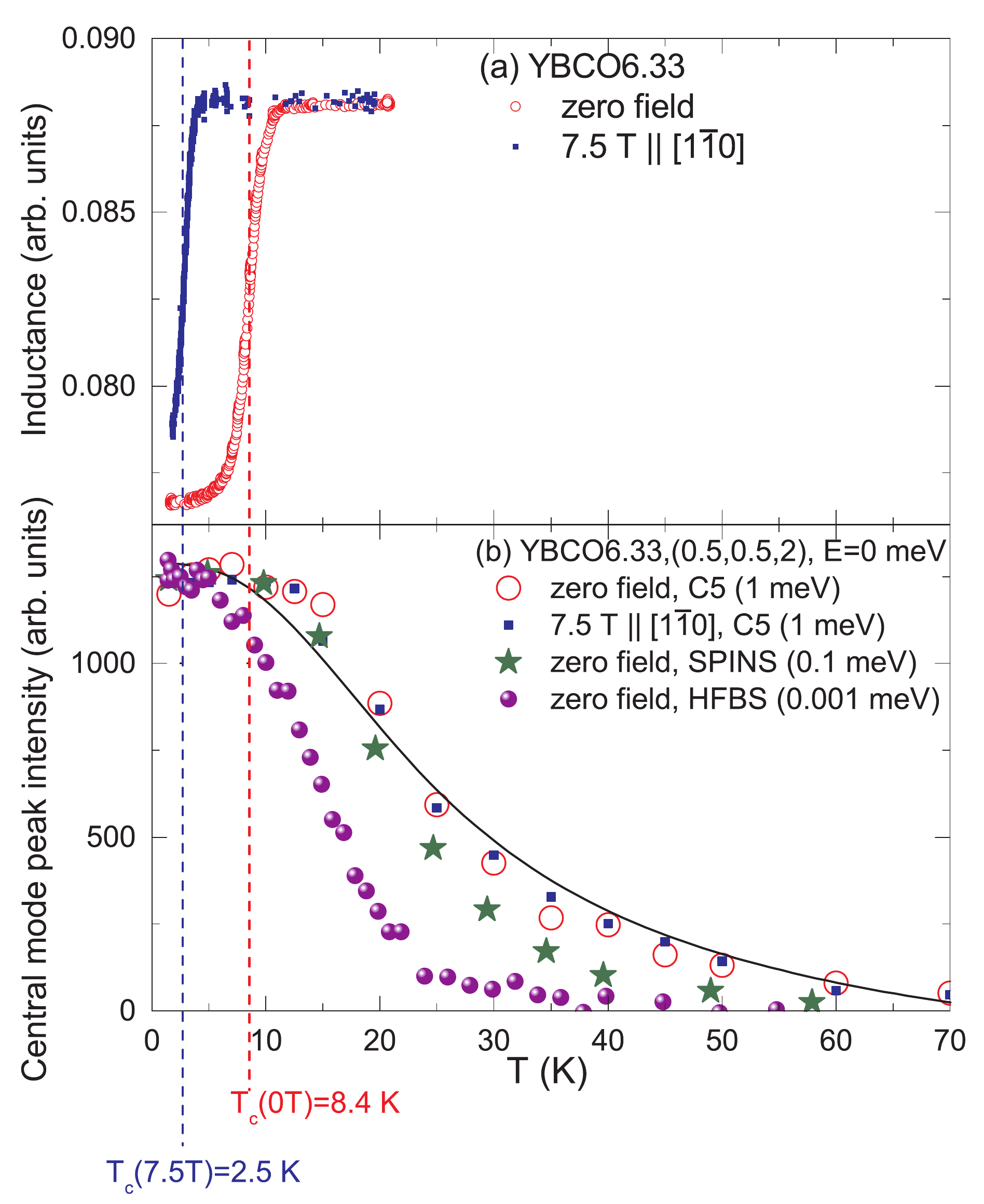}}\vskip.2cm \caption []{(a) The superconducting transition is observed at 8.4 K at zero field and is reduced to 2.5 K at 7.5 T $||$ [1$\overline{1}$0]. The sample was located in a coil and the inductance of the coil, $L$, at 17 Hz was monitored as a function of temperature using an inductance bridge. The inductance of the coil is related to the susceptibility of the sample, $\chi_S$(T), through $L=L_0(1 + 4\pi \chi_S$(T)). When the sample becomes superconducting, $\chi_S$(T) becomes diamagnetic and hence the inductance of the coil will decrease. (b) Central mode intensity at E=0 at $\mathbf{Q}_{\mathrm{AF}}$=(0.5 0.5 2) at zero field and at 7.5 T $||$ [1$\overline{1}$0] observed with thermal neutrons (resolution$\sim$1 meV). Despite the change in T$_c$, the central mode is not affected by the magnetic field. Also shown is the intensity at zero field observed with cold neutrons (resolution$\sim$0.1 meV) and with backscattering (resolution$\sim$1 $\mu$eV). The intensity observed for each measurement is normalized to its maximum at low temperatures. The temperature scale of the central mode intensity decreases when measured with better energy resolution. The solid line is a guide to the eye.  
} \label{fig2}
\end{center}
\end{figure}

The experiments were performed on large single crystals prepared at University of British Columbia (UBC) by a top-seeded melt growth technique~\cite{peets02}. The crystals each of volume $\sim$1 cm$^3$ were well annealed and in the orthorhombic phase with room temperature lattice parameters $a$=3.844 \AA, $b$=3.870 \AA~(for data analysis purposes we take $a$=$b$=$(a+b)/2=$3.857 \AA), and $c$=11.791 \AA. The superconducting transition temperature, T$_c$, was determined to be 5 K after the crystal growth and rising to 8.4 K after further annealing for several months at room temperature at Chalk River. The superconducting transition temperature determined by AC Meissner susceptibility in a coil at 17 Hz remains very sharp for such a low doping (the transition width is only 2 K), see Fig.~\ref{fig2}(a). T$_c$ is reduced to 2.5 K in a magnetic field of 7.5 T $||$ [1$\overline{1}$0]. The sample hole doping $p$=0.055 was determined from the observed T$_c$ in zero field and the empirical formula~\cite{tallon95}: 1-T$_c$/T$_{\mathrm{max}}$=82.6[$p$-0.16]$^2$ with T$_{\mathrm{max}}$=94.3 K which is also consistent with the c-axis lattice constant vs. doping relation observed~\cite{liang06} for YBCO6+x. For the zero-field experiments, four crystals (4 cc in total with a mosaic spread of 2$^\circ$) were co-aligned in the (HHL) plane. For the experiment in an applied magnetic field, one crystal (1 cc with a mosaic spread of 1$^\circ$) was used with the field applied along [1$\overline{1}$0].

Thermal neutron scattering experiments were made with the C5 spectrometer at NRU reactor at Chalk River Laboratories with pyrolytic graphite PG(002) crystals vertically focusing for the monochromator and flat for the analyzer with a fixed final energy of E$_f$=14.6 meV. Elastic and inelastic measurements of the sample aligned in the (HHL) plane were made in the temperature range 1.5 K to 300 K, in magnetic field up to 7.5 T, up to 16 meV energy transfer in the (0.5 0.5 2) zone and up to about 40 meV in the (0.5 0.5 5) zone. Fast neutrons were removed from the incident beam by a liquid nitrogen cooled sapphire filter before the monochromator. Two PG filters with a total thickness of 10 cm in the scattered beam removed higher-order neutron wavelengths. The horizontal collimations were controlled with Soller slits as specified in the plots of data. A search revealed no oxygen superlattice peaks indicating that any oxygen ordering (i.e. Ortho-II and/or Ortho-III) in the CuO chains is short-ranged. Nonetheless a small a-b splitting was detected with x-rays, showing that the sample remains in the orthorhombic phase.

Cold neutron measurements were performed with the SPINS spectrometer at the NIST Center for Neutron Research (NCNR) and with PG(002) crystals vertically focusing for the monochromator and flat for the analyzer with a fixed final energy of E$_f$=2.9 meV and a collimation of [guide 80$'$ S 80$'$ 80$'$]. For energy transfers greater than 2 meV a cooled Be filter was placed in the scattered beam. For energy transfers less than 2 meV a cooled Be filter was placed before the sample. Data was normalized from results at overlapping energies.

To further investigate how the temperature dependence of magnetic correlations depends on the experimental energy resolution, we also carried out neutron scattering measurements with the NCNR's high-flux backscattering spectrometer, HFBS, with an energy resolution of $\sim$1 $\mu$eV. Elastic scans were performed at a fixed final energy of E$_f$=2.08 meV using a Si(111) monochromator mounted on a Doppler drive. Comparison of the temperature dependence of the observed magnetic elastic (central mode) data at (0.5 0.5 2), with thermal, cold and backscattered neutrons is shown in Fig.~\ref{fig2}(b). 

The magnetic scattering function $S(\mathbf{Q},\omega)$, for momentum transfer $\mathbf{Q}$, and energy transfer $\omega$, can be determined directly from the measured magnetic neutron scattering intensity by
\begin{equation}\label{magsection}
   I(\mathbf{Q},\omega)~\propto~f^2(Q) B^2(Q_c) S(\mathbf{Q},\omega)
\end{equation}
\noindent where $f(Q)$ is the anisotropic Cu$^{2+}$ form factor~\cite{shamoto93} and $B(Q_c)=2 \mathrm{sin}( \frac{c Q_c }{2} z')$ is the independent bilayer structure factor with $Q_c=2\pi L/c$ and $z'=1-2z_\mathrm{{Cu2}}$ (see the Appendix). The model used for $S(\mathbf{Q},\omega)$ in the data analysis is discussed later in Section~\ref{results}. The measured data is fitted to a function which is the convolution~\cite{reslib} of this model with the spectrometer resolution function to obtain the model parameters. All uncertainties and error bars are standard errors. 

The inelastic data are corrected for contamination of the incident beam monitor by higher wavelength harmonics~\cite{shiraneBook}. A transverse acoustic phonon measured close to the (006) Bragg peak is used to put the observed intensities on an absolute scale, see Appendix. The total elastic magnetic moment is obtained by integrating over the observed elastic line in absolute units in energy and in momentum transfer along [HH0] and [00L] directions relative to the magnetic Brillouin zone centre. The results are consistent with the magnetic moment derived from the analysis of several nuclear Bragg peaks.

\begin{figure}
\vskip 0cm
\resizebox{1\linewidth}{!}{\includegraphics{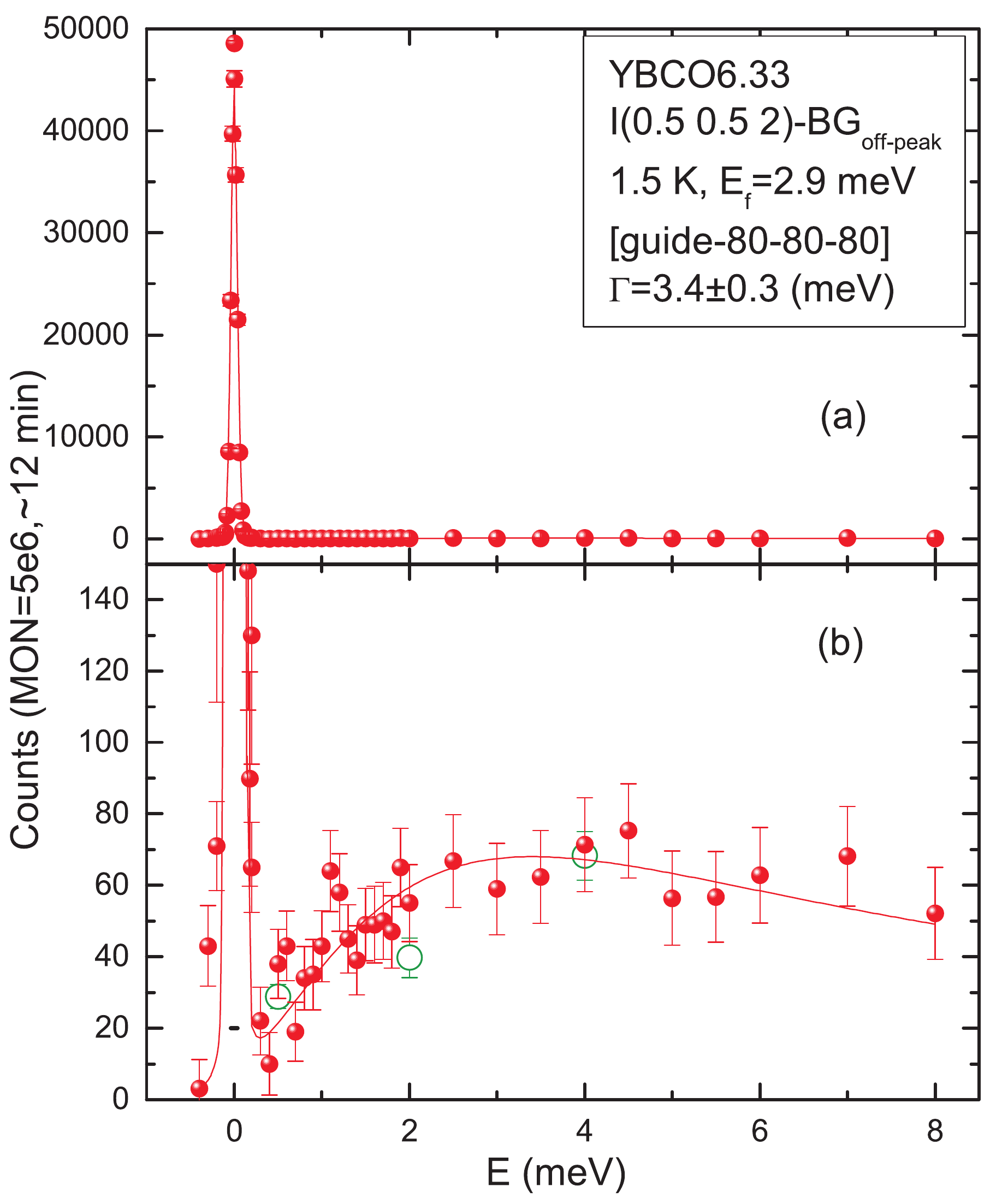}}\vskip
0cm \caption []{Inelastic spectrum measured with cold neutrons at the AF center $\mathbf{Q}_{\mathrm{AF}}$=(0.5 0.5 2) at 1.5 K. The average of the intensity at (0.3 0.3 2) and (0.7 0.7 2) is used as background (BG$_{\textrm{off-peak}}$). Scattering comprises an intense resolution limited peak centered at zero energy transfer (a), and a much weaker broad component as seen in panel (b). The solid line is a fit to a resolution-limited Lorentzian at zero energy and to a broad relaxational Lorentzian with a $\sim$3.4$\pm$0.3 meV relaxation rate. The horizontal line in (b) is the energy resolution at zero energy transfer. The open circles in (b) are the peak height of the fitted constant energy H-scans along (0.5 0.5 2), and they agree with two-point H=0.3 and 0.7 background subtraction.  } \label{fig3}
\end{figure}


\section{Results}\label{results}

\subsection{Two energy scales}

The magnetic spectrum at the antiferromagnetic (AF) center $\mathbf{Q}_{\mathrm{AF}}$=(0.5 0.5 2) at 1.5 K, measured with cold neutrons with an energy resolution of 0.08 meV, is shown in Fig.~\ref{fig3}. The average of the scattering observed at (0.3 0.3 2) and (0.7 0.7 2) is used as background (BG$_{\textrm{off-peak}}$). This method of background subtraction is verified by constant-energy Q-scans at several energy transfers (see open circles Fig.~\ref{fig3} (b)). The spectrum exhibits two energy scales: a very slow or static response characterized by a resolution limited elastic peak at zero energy (central mode) and a much weaker and broader spectral feature. All spectra measured such as Fig.~\ref{fig3}(b) indicate that there is no spin gap. At low energies the inelastic magnetic scattering linearly decreases to zero with decreasing energy as may be expected from overdamped spin waves (paramagnons). The observed absence of dynamic spin spectral weight as energy goes to zero can only help formation of superconducting pairs for T$<<$T$_c$. Any N\'{e}el order would instead have led to a growth in spin spectral weight as energy goes to zero, much as upon approach to a 3D N\'{e}el phase.

In the following we present the temperature and wavevector dependence of the central mode and the inelastic feature. Our results demonstrates that the central mode represents only short-range magnetic correlations and that the dynamic overdamped fluctuations give rise to the broad inelastic feature. 

\begin{figure*}[tbh]
\vskip 0cm
\resizebox{.9\linewidth}{!}{\includegraphics{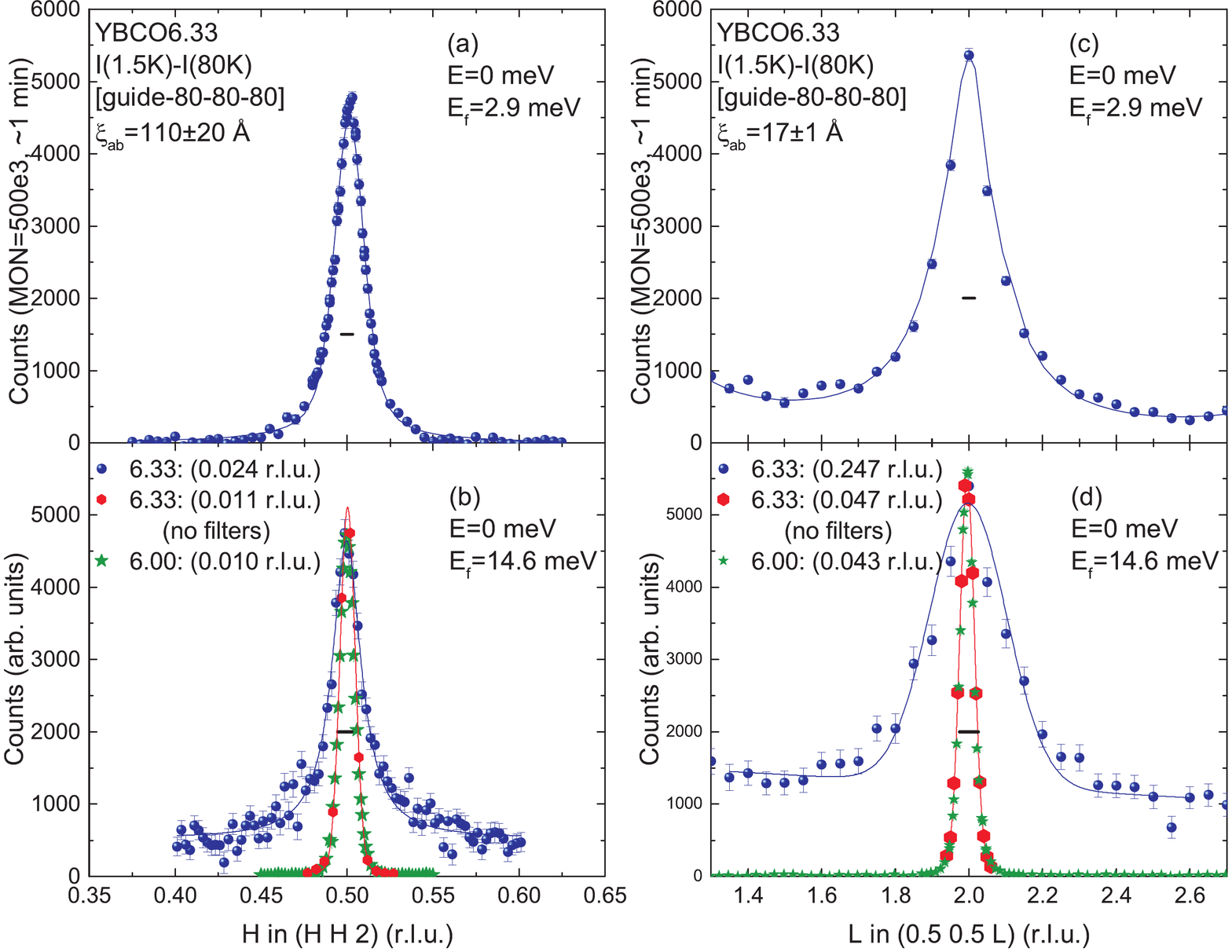}}\vskip 0cm \caption []{Spin correlations from elastic scans along the [HH0] and [00L] through the AF center $\mathbf{Q}_{\mathrm{AF}}$=(0.5 0.5 2). Data collected with cold neutrons (E$_f$=2.9 meV) at 1.5 K are shown in panels (a) and (c) and with thermal neutrons (E$_f$=14.6 meV) at 3 K in panels (b) and (d). The horizontal lines are the calculated resolution Bragg widths (FWHM). To make a direct comparison with the resolution of the spectrometer at this momentum transfer, we removed the PG filters and repeated the scans (red hexagons). In addition, we measured the magnetic scattering from an AF Bragg ordered single crystal of YBCO6.00 with a transition to a long-range AF order (green stars). The magnetic scattering for YBCO6.33 is wider than both the measured resolution of the spectrometer and the magnetic Bragg peak observed for YBCO6.00. Gaussian FWHM for each scan is given in brackets in the panels b and d. This indicates that only short-ranged magnetic correlations exist in YBCO6.33. The fits are obtained by convolution of the 4D resolution with an elastic Lorentzian function with isotropic q-width in the a*-b* plane and a larger width along c*.  
} \label{fig4}
\end{figure*}

\subsection{Quasielastic scattering: central mode}

\begin{figure}[tbh!]
\begin{center}
\resizebox{.87\linewidth}{!}{\includegraphics{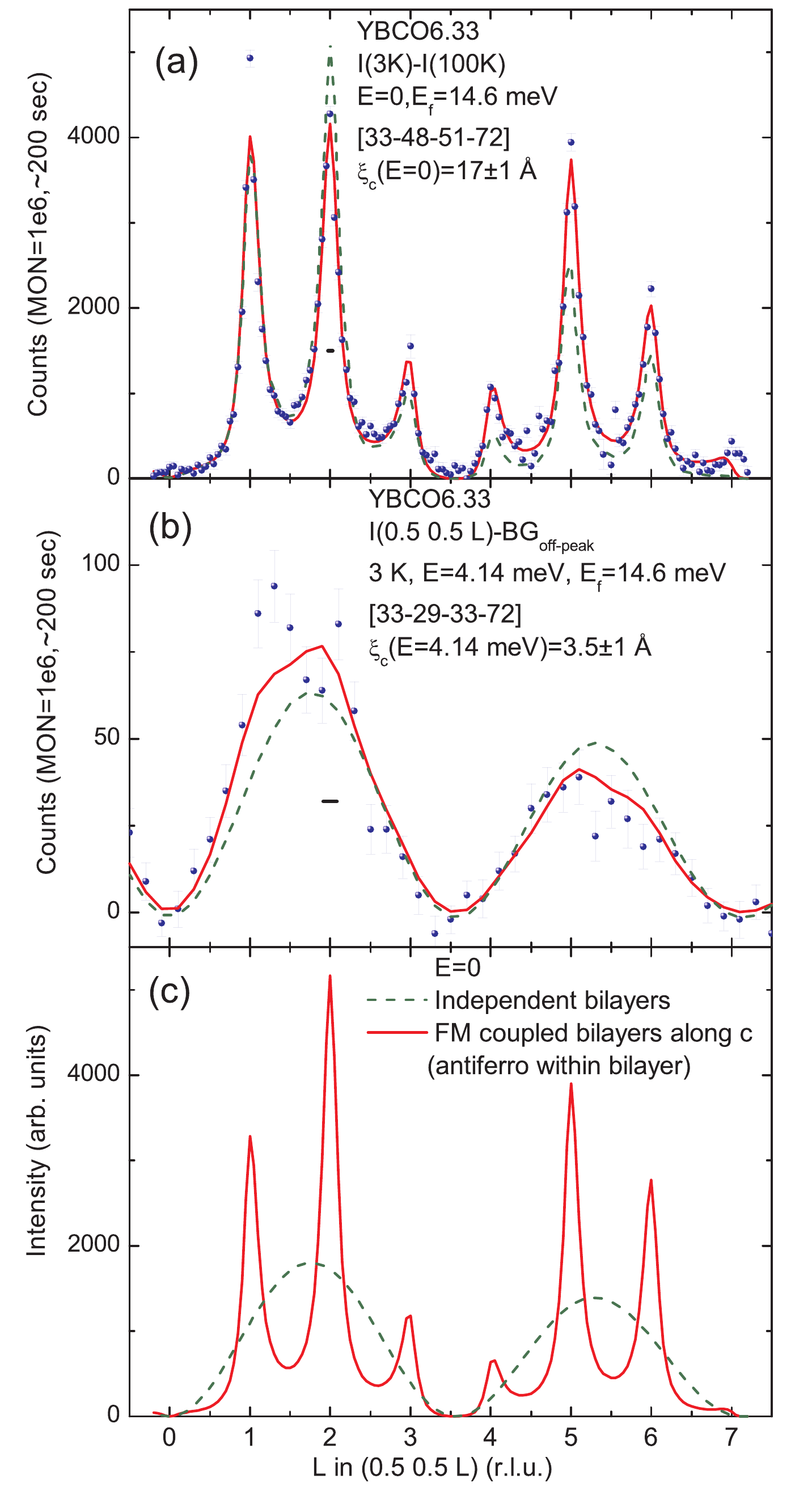}}\vskip -.12cm \caption []{(a) Approach to 3D via out-of-plane elastic spin correlations along [00L] through [0.5 0.5 L] at 3 K where data at 100 K is used as background. Fits are shown of resolution-convoluted Lorentzians centered at integer L (solid red line) and to the coupled bilayer model (dashed green line). For the convolution with resolution the in-plane correlation lengths are taken from elastic scans along [H H 0] through (0.5 0.5 2), similar to the data shown in Fig.~\ref{fig4}. (b) The L-scan at a finite E=4.14 meV energy transfer shows that spin excitations mainly originate from independent bilayers (dashed green line). The data is corrected for background taken as the average of the observed intensity at (0.3 0.3 L) and (0.7 0.7 L), BG$_{\textrm{off-peak}}$. (c) A simulation of scattering along the L-direction for uncoupled bilayers (dashed green line) and ferromagnetically coupled bilayers (solid red line). Horizontal lines at L=2 in (a) and (b) give the calculated resolution widths in L (FWHM) at E=0 and 4.14 meV. In (a) and (c) the incipient 3D pattern is larger than in (b) because the static susceptibility is much larger as the energy approaches zero.} \label{fig5}
\end{center}
\end{figure}

Figs.~\ref{fig4} and~\ref{fig5}(a) show typical elastic Q-scans through (0.5 0.5 2) along [HH0] and [00L] directions observed at base temperature with a background subtracted at 80 K and 100 K respectively. The lack of temperature dependence above $\sim$80 K justifies attribution of the growing scattering at low temperatures to the magnetic scattering. This is confirmed by polarized neutron scattering~\cite{yamanipol}. 

The scattering appears in form of a peak centered at the AF wavevector, $\mathbf{Q}_{\mathrm{AF}}$ =(0.5 0.5 L) with L=integer and broader than the resolution. To make a comparison of the observed peak widths with the resolution of the spectrometer, we directly measured the resolution by removing the PG filters and repeating the exact scans. The calculated resolution is in good agreement with the measurement (horizontal lines) as seen in Figs.~\ref{fig4}(b) and (d). In addition, we made a comparison with the observed scattering from an undoped YBCO6.00 crystal with a N\'{e}el transition to a long range antiferromagnetic state at  $\sim$430 K. We find that the peak widths for the undoped crystal are similar to the resolution and smaller than the ones in YBCO6.33 along both directions. This comparison confirms in this low doped YBCO6.33 superconductor the correlation lengths of the spins remain finite and long-ranged AF order is absent. The fact that the scattering is peaked at integer L-values, however, indicates the 3D spin correlations are parallel (ferromagnetic) between bilayers in adjacent cells along the c-axis. The spins in the two planes of a bilayer in one cell always remain antiparallel.

The magnetic scattering function in Eq.~\ref{magsection}, $S(\mathbf{Q},\omega)$, can be written as:

\begin{equation}\label{magscatfun}
   S(\mathbf{Q},\omega)=|F(\mathbf{Q})|^2  ~ g(\omega)
\end{equation}

\noindent where $|F(\mathbf{Q})|^2$ and $g(\omega)$ describe the momentum and spectral forms of the scattering, respectively. To extract the zero-frequency correlation lengths for both reciprocal space directions, we fitted the elastic data along [HH0] and [00L] to the convolution of the four-dimensional resolution function~\cite{reslib} with the elastic magnetic scattering function of the form,
\begin{equation}\label{magscatfunelastic}
   S(\mathbf{Q},\omega \simeq 0)=|F(\mathbf{Q})|^2  ~\delta(\omega)
    \end{equation}

\noindent To avoid problems with the convolution, the delta function was slightly broadened to
\begin{equation}\label{deltafunc}
\delta(\omega)=\frac{\gamma}{\omega^2+\gamma^2}
    \end{equation} 

\noindent with $\gamma \ll \Delta E_\mathrm{res}$ fixed where $\Delta E_\mathrm{res}$ is the energy resolution, with $\gamma$ typically of order 1 $\mu$eV and 10 $\mu$eV for cold and thermal neutron data sets, respectively.

We first modeled the momentum dependent spectral form, $|F(\mathbf{Q})|^2$, with a 
a set of independent Lorentzians, 
\begin{equation}\label{QdepLor}
|F(\mathbf{Q})|^2=~\frac{A_{CM}}{1+q_{ab}^2 \xi_{ab}^2} ~\sum_i  \frac{A_{i}}{1+q_{c,i}^2 \xi_{c}^2}
\end{equation}
where $q_{ab}=(H-0.5, K-0.5)(\frac{2\pi}{a})$, $q_{c,i}=(L-L_i)(\frac{2\pi}{c})$ with $L_i$ an integer ranging from -1 to 7, and $\xi_{ab}$ and $\xi_{c}$ are correlation lengths in the ab-plane (assumed symmetric along the a and b in-plane directions) and along the c-axis, respectively. These Lorentzians describe the 3D correlations along L and their relative intensities, $A_{i}$, are allowed to vary independently. Because fitting of $\xi_{ab}$ to an H-scan involves resolution convolution over a known L width ($\xi_c$), and vice versa, we iterated to ensure that a consistent set of $\xi_{ab}$ and $\xi_c$ were fitted to the two data scans (Figs.~\ref{fig4} and ~\ref{fig5}(a)). We extract the intrinsic low temperature elastic spin correlation lengths at 1.5 K to be $\xi_{ab}$=110$\pm$20 \AA\ in the ab-plane and $\xi_{c}$=17$\pm$1 \AA\ for all peaks along the c-direction. Both cold and thermal data gave the same values within the error bars after the resolution effects are included. Directly from the observed widths of the data along [HH0] and along [00L] directions a correlation length of about 40-50 \AA\ is obtained for in-plane correlations  ($1/ \xi_{ab}=\sqrt2\Delta \mathrm{H} (2\pi/a$)) and about 11-17 \AA\ for correlation length along the c-direction ($1/ \xi_{c}=\Delta \mathrm{L}(2\pi/c))$, where $\Delta \mathrm{H}$ and $\Delta \mathrm{L}$ are the observed Lorentzian HWHM in r.l.u. along H and L, respectively.

We may also extract the c-axis correlation length, $\xi_{c}$, by assuming that the bilayer amplitude decays exponentially along the c-axis as $\mathrm{exp}(-r_z/\xi_{c})$. The Fourier transform of this correlation is exactly given~\cite{yoshizawa79} by

\begin{equation}\label{Qdepbilayer}
   |F(\mathbf{Q})|^2=~\frac{A_{CM}}{1+q_{ab}^2 \xi_{ab}^2} ~\frac{\mathrm{sinh}(\xi_{c}^{-1} c/2)~\mathrm{cosh}(\xi_{c}^{-1} c/2)}{\mathrm{sinh}(\xi_{c}^{-1} c/2)^2+\mathrm{sin}(Q_c c/2)^2}
    \end{equation}
For large $\xi_{c}$, Eq.~\ref{Qdepbilayer} gives a set of Lorentzians along L centered at each integral $L_i$ but with a single $\xi_{c}$ (of the form given by Eq.~\ref{QdepLor}). This coupled bilayer formula gives the intensity of the peaks along L through the bilayer structure factor $B^2(Q_c)$ and eq.~\ref{magsection}. The $\xi_{c}$ =16 $\pm$ 1 {\AA} extracted from a fit of the coupled bilayer model agrees with the value obtained using the independent Lorentzians of Eq. (5). The coupled bilayer model used in Ref.~\onlinecite{stock06} is the square of Eq.~\ref{Qdepbilayer} and hence the c-axis correlation lengths reported there correspond to a Lorentzian-squared model.

Fig.~\ref{fig5}(a) shows that the coupled bilayer model describes the data almost as well as the independent-Lorentzian model but underestimates the intensity ratio of the peak at L=1 to that at L=2. Since the intensity recovered somewhat when  cold neutrons were used, we attribute the difference to unknown resolution effects. The fit is equally good to a Lorentzian squared with a correlation length along the c-axis of $\sim$7.8$\pm$0.5 \AA, shorter as expected.

\begin{figure*}[tbh]
\begin{center}
\vskip 0cm
\resizebox{1\linewidth}{!}{\includegraphics{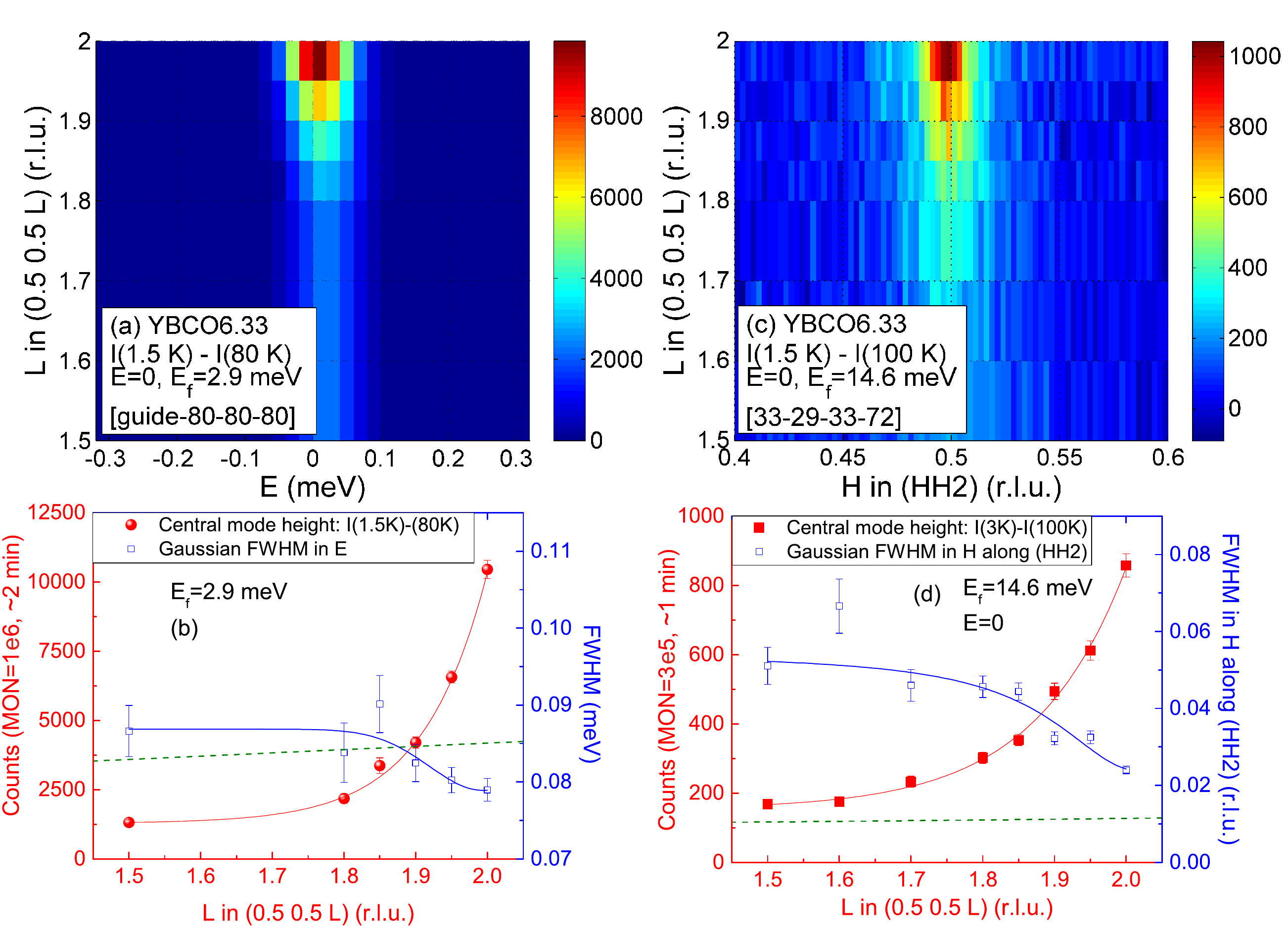}}\vskip 0cm \caption []{(a) Quasielastic energy scans of the central mode as a function of L close to AF center $\mathbf{Q}_{\mathrm{AF}}$=(0.5 0.5 2) at 1.5 K. (b) Central mode peak height and energy FWHM obtained from scans in (a) by fitting to a Gaussian function. (c) Elastic H-scans of the central mode as a function of L at 3 K. (d) Central mode peak height and FWHM in H obtained from (HH2) scans in (c). In (b) and (d) dashed lines are the calculated resolution and the solid lines are guides to the eye.} \label{fig6}
\end{center}
\end{figure*}


In Fig.~\ref{fig6} we show the properties of the central mode in the L-E and L-H planes. From Gaussian fits we plot in panels (b) and (d) the peak height, and the energy and momentum widths. As L$\rightarrow$2, the central mode amplitude grows while the relaxation rate slows to the resolution limit of 0.08 meV. Right at the integer-L momentum, the spins are frozen on the neutron timescale. The smooth behaviour in peak height, energy and momentum width as L tends to the 3D point L=2 shows that there is no Bragg feature hidden within the central mode correlations. Thus the spin correlations are short range and no N\'{e}el order is present. The near-critical scattering does not follow a Lorentzian form, for the width in H would then continue to grow as L departed from L=2. Instead the H width becomes constant far from L=2. This behaviour shows that the quasielastic scattering far from the L-integral peaks adopts a rod-like form typical of 2D systems with a constant width in H.

\begin{figure*}[tbh]
\begin{center}
\vskip 0cm
\resizebox{.95\linewidth}{!}{\includegraphics{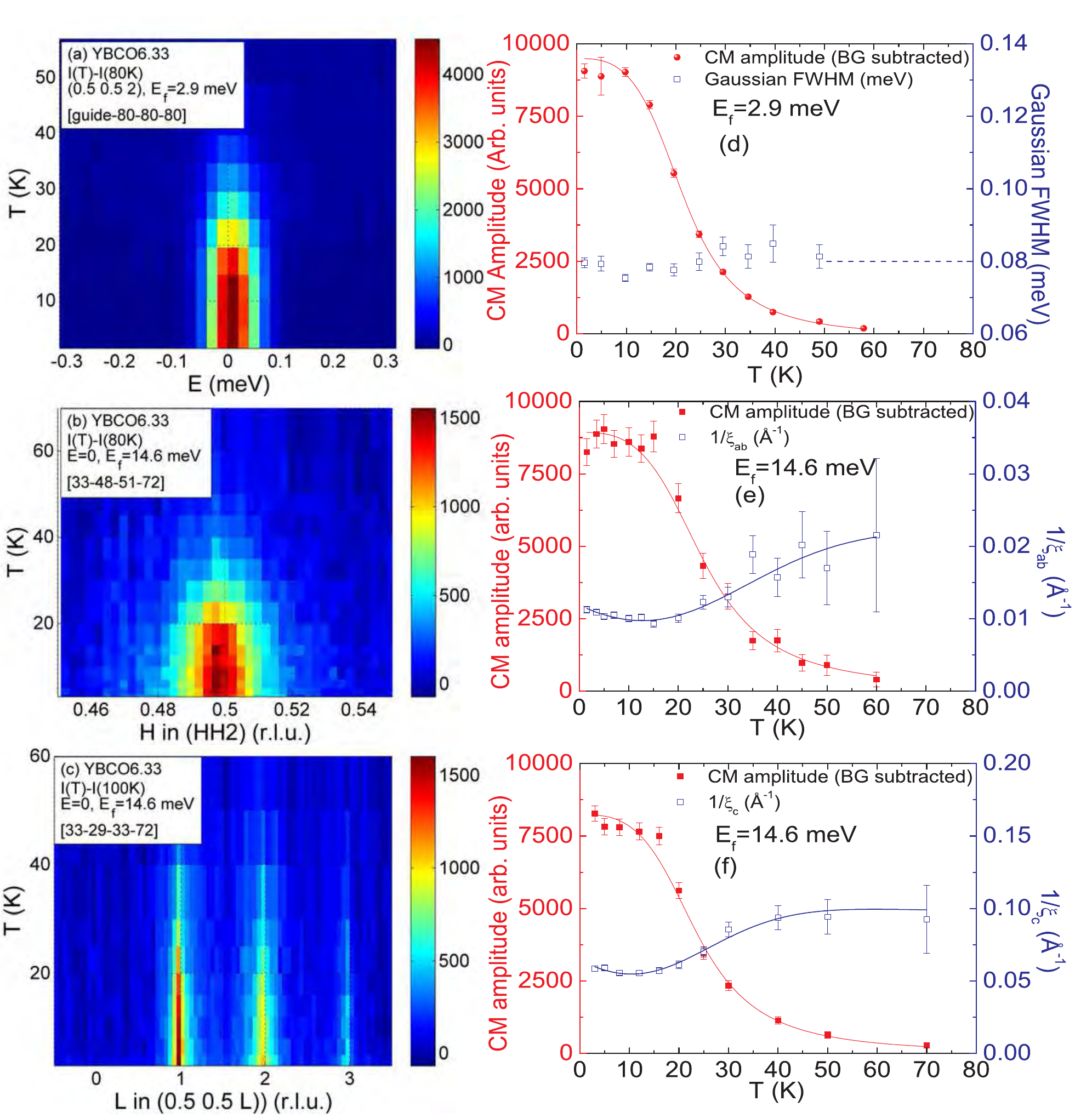}}\vskip .0cm \caption []{ The temperature dependence of the intensity around $\mathbf{Q}_{\mathrm{AF}}$ =(0.5 0.5 2) in (a) energy scans around zero energy transfer, (b) in elastic (HH0) scans and (c) in elastic (00L) scans through (0.5 0.5 2). The data is corrected for background by subtracting the data at 80 K in (a) and (b) and at 100 K in (c).  The dotted horizontal line in (d) is the calculated resolution width (FWHM). No signature of a N\'{e}el transition down to 1.5 K is found in the temperature dependence of the central mode amplitude. The central peak is consistent with being resolution limited in energy at low temperatures. As seen in panels (e) and (f), the spatial correlation lengths in the ab-plane and along the c-direction lengthen gradually on cooling but remain finite at the lowest temperatures.  }\label{fig7}
\end{center}
\end{figure*}

Figs.~\ref{fig7}(a,b,c) show the temperature dependence of the quasielastic scattering near (0.5 0.5 2) by scans in energy, H and L. The fit parameters including central mode amplitude, the quasielastic energy width FWHM obtained from a gaussian fit, and the inverse correlation lengths obtained from the convolution of resolution with Eq.~\ref{QdepLor} are shown in Figs.~\ref{fig7}(d,e,f). We find that the quasielastic amplitude grows monotonically on cooling before saturating below about 15 K. There is no anomaly that might signal a transition to long-range AF order. In addition, the transition to the SC state at T$_c$=8.4 K does not affect the temperature dependence of the central mode indicating that superconductivity and AF spin coupling develop independently. 

The inverse correlation lengths along both directions, $\xi_{ab}$ and $\xi_{c}$, remain finite at all temperatures. The $\xi_{ab}$ gradually decrease on cooling while $\xi_{c}$ exhibits a more pronounced decrease below $\sim$30 K. The spins are attempting to find a state similar to the AF state below 30 K but are prevented by hole doping. Both inverse correlation lengths may hint at the presence of a minimum around $\sim$15 K. As well the central mode dynamics slows on cooling until reaching the resolution limit. A similar narrowing of the dynamic width was observed~\cite{stock06} in YBCO6.35 (T$_c$=18 K) but the increase of the spin correlation lengths below 15 K on cooling in YBCO6.33 (T$_c$=8.4 K) is a new effect (Figs.~\ref{fig7}(e,f)). The growth of the central mode amplitude cannot be accounted for by the decrease in its extent in momentum-energy space, $\xi_{ab}^{-2}~ \xi_{c}^{-1} ~\Delta E$, all of which decrease from 80 K to 15 K.

A comparison of the temperature dependence of the central mode intensity measured with different resolutions shown in Fig.~\ref{fig2}(b) indicates that the temperature scale of the central mode depends on the energy resolution at which the measurement is performed. When the central mode is probed with high-resolution backscattering with an energy resolution of only $\sim$1 $\mu$eV, the scattering appears on cooling only below $\sim$40 K compared to $\sim$80 K for thermal neutrons with broader energy resolution of $\sim$1 meV. 
Hence all thermal and cold neutron measurements are extremely resolution limited, e.g., resolution of 0.08 meV is 80 times wider than the maximum intrinsic width of 1 $\mu$eV of the central mode. The dependence of the central mode intensity vs. T on resolution is similar to the behaviour observed in spin glasses at low temperatures~\cite{murani78,aeppli85,matsuda05}.

\begin{figure*}[tbh]
	\begin{center}
		\vskip 0cm
		\resizebox{.9\linewidth}{!}{\includegraphics{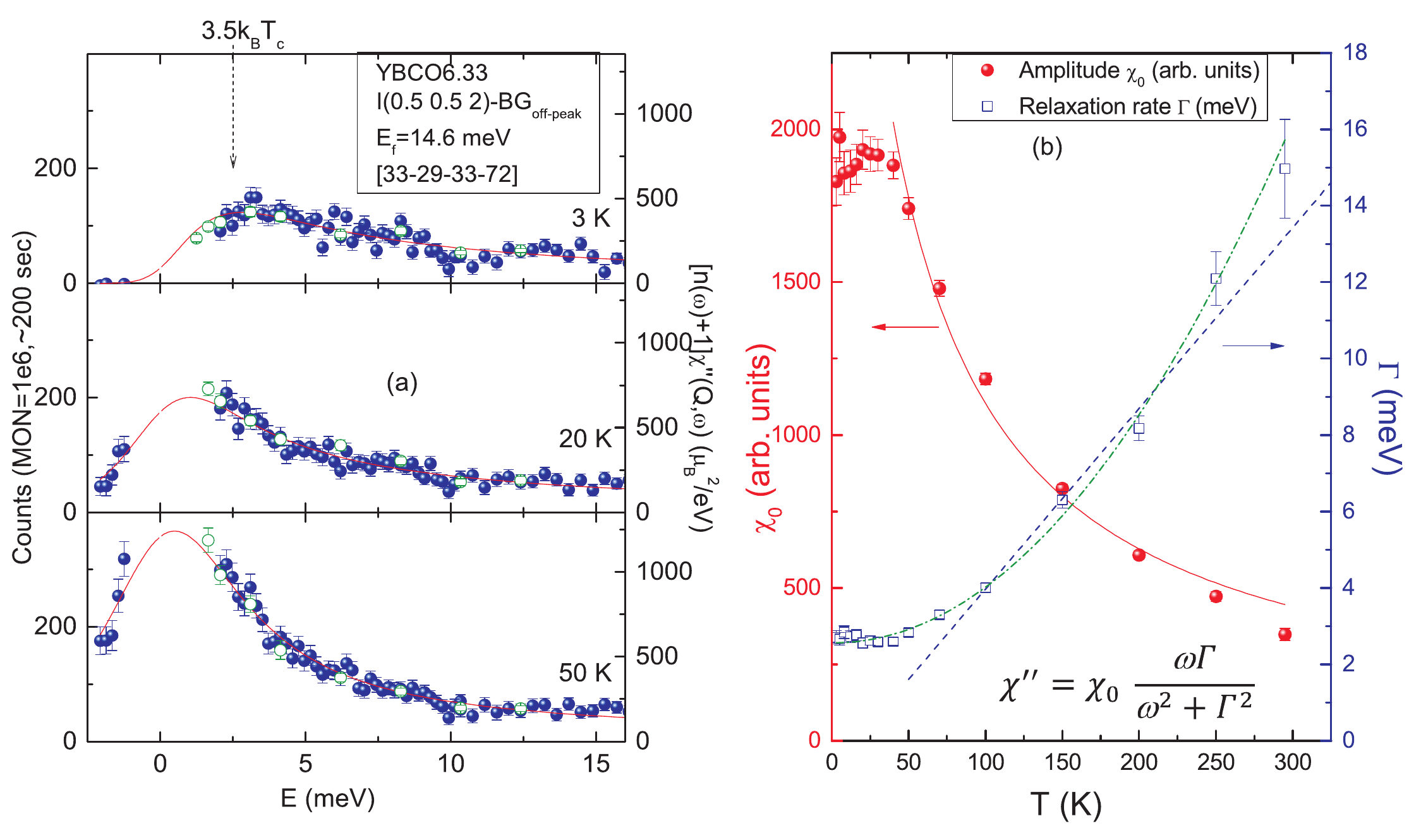}}\vskip
		0cm \caption []{  (a) Temperature dependence of inelastic spectra observed at the AF center $\mathbf{Q}_{\mathrm{AF}}$= (0.5 0.5 2). At each temperature the averaged scattering measured at (0.3 0.3 2) and (0.7 0.7 2), BG$_{\textrm{off-peak}}$, is used for the background subtraction (solid circles). This method of background subtraction is confirmed by constant-energy (HH0) scans through the AF center $\mathbf{Q}_{\mathrm{AF}}$=(0.5 0.5 2) resulting in the same intensity (open circles). The solid lines are fits to a broad damped response with a relaxation rate, $\Gamma$, given by Eq.~\ref{modlor}. Note that for these fits, the data within the elastic resolution region is removed. (b) The temperature dependence of the model parameters $\chi_0$=$A_{ex}$ and $\Gamma$. A Curie-Weiss high temperature trend for $\Gamma$ is shown by a dashed line, and that for the static susceptibility by a solid line. We find that a quadratic temperature dependence describes the relaxation rate, $\Gamma$(T)=a + bT$^2$, over the entire temperature range (the dash-dotted line).  }
		\label{fig8}
	\end{center}
\end{figure*}

\subsection{Inelastic spectra}

Typical inelastic spectra measured by constant-Q energy scans at (0.5 0.5 2) are shown in Fig.~\ref{fig8}(a) at different temperatures. The spectra are corrected for the wavelength feedthrough in the monitor~\cite{shiraneBook}. The average of the scattering at (0.3 0.3 2) and (0.7 0.7 2), BG$_{\textrm{off-peak}}$, is used for background subtraction. This method of background subtraction is confirmed by constant-energy (HH2) scans through the AF center $\mathbf{Q}_{\mathrm{AF}}$=(0.5 0.5 2) resulting in the same peak intensity (open circles in Fig.~\ref{fig8}(a)). We have used the phonon calibration method (see Appendix) to put the observed scattering on an absolute scale (right-hand axis in Fig.~\ref{fig8}(a)). 

The background corrected data is fitted to the convolution of the 4D resolution function with the model of Eq.~\ref{magscatfun} where $g(\omega)$ is given by

\begin{equation}\label{modlor}
g(\omega)=~ \frac{\gamma A_{CM}}{\omega^2
+\gamma^2} + [1+n(\omega)]\frac{\omega \Gamma A_{ex}}{\omega^2 +\Gamma^2}
    \end{equation}
\noindent where $[1+\mathrm{n}(\omega)]$=$1/[1-\mathrm{exp}(-\hbar\omega/k_BT)]$ is the Bose population factor. The first term describes an energy resolution-limited central peak forced to be elastic by setting $\gamma \ll \Delta E_\mathrm{res}$. The second term represents spin diffusion described by a modified Lorentzian defined by its relaxation rate, $\Gamma$. The concurrent Lorentzian momentum dependence (Eq.~\ref{QdepLor}) displays a memory loss over a time ($1/\Gamma$) and over long but finite corresponding correlation lengths in the ab-plane and very short distances along the c direction. The solid lines in Fig.~\ref{fig8}(a) show that the model can fit the data reasonably up to 10--15 meV. We find that at higher energy transfers the data falls less rapidly than the model. Extra scattering at energies larger than 15 meV could form a precursor to the almost constant local susceptibility expected for 2D
spin waves at larger energies. At larger energies, E$\sim$20 meV, phonon contributions are known~\cite{stock04} to occur. Note that the width in q is constant for E$<$10 meV (Fig.~\ref{fig11}(b)) so that the spectrum at $\mathbf{Q}_{\mathrm{AF}}$=(0.5 0.5 2) is proportional to the local susceptibility.


\begin{figure}[tbh!]
\begin{center}
\vskip 0cm
\resizebox{1\linewidth}{!}{\includegraphics{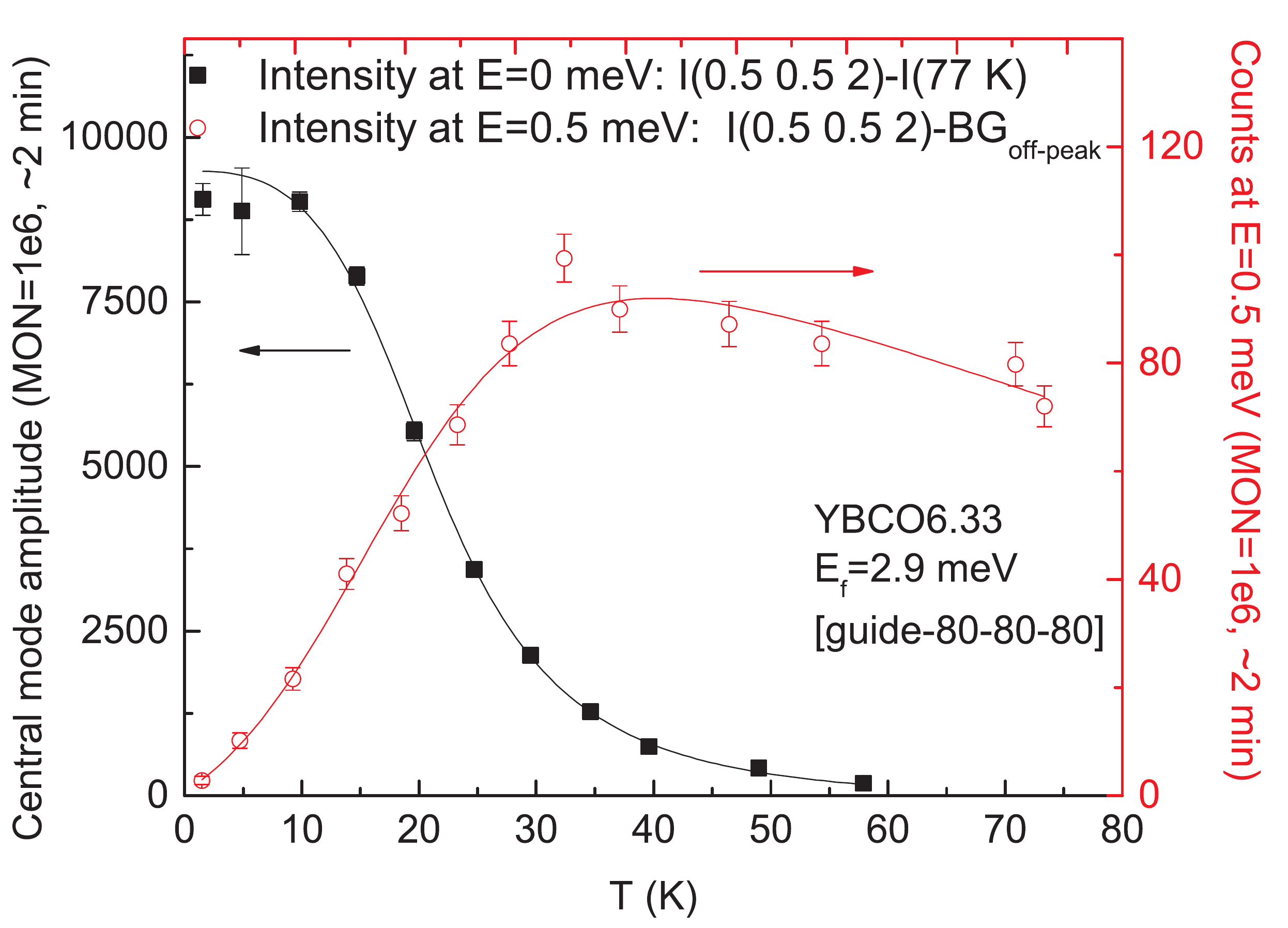}}\vskip
0cm \caption []{A comparison at (0.5 0.5 2) of the intensity of the central mode with that of the low energy fluctuations at E=0.5 meV as a function of temperature. A spectral transfer from dynamic fluctuations to elastic correlations is observed as the system is cooled down to its ground state. Solid lines are guides to the eye. The characteristic temperature of $\sim$30 K for the depopulation of the fluctuations corresponds to the relaxation rate of $\sim$3 meV and not to the 0.5 meV fluctuation energy.  The loss of fluctuation strength appears in the growth of the ground state represented by the central mode.} \label{fig9}
\end{center}
\end{figure}

\begin{figure}[tbh!]
	\begin{center}
		\vskip 0cm
		\resizebox{.95\linewidth}{!}{\includegraphics{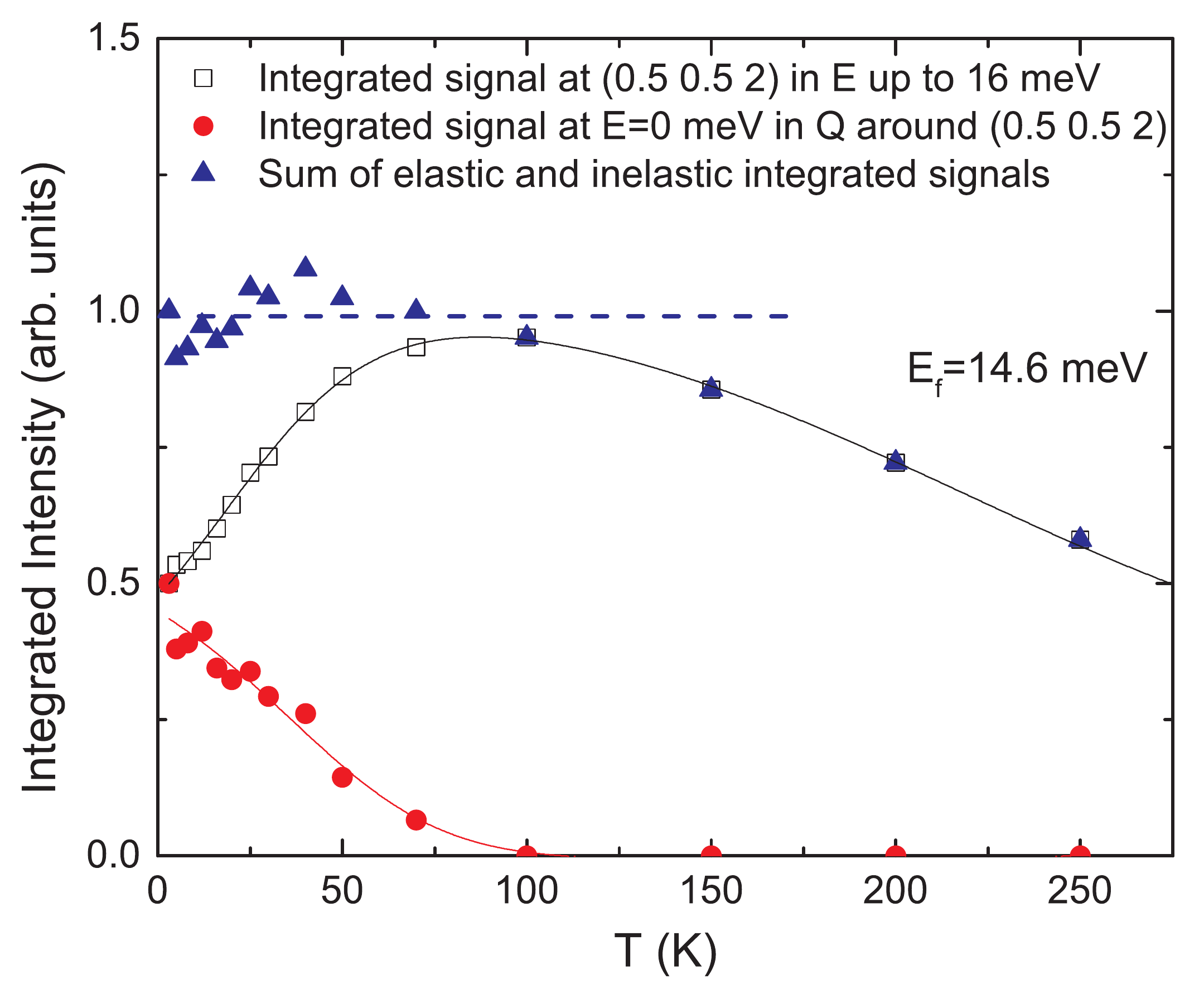}}\vskip
		0cm \caption []{Temperature dependence of the integrated intensity of elastic and inelastic scattering. As the inelastic scattering decreases on cooling, the elastic component increases. Solid and dashed lines are guides to the eye.  } \label{fig10}
	\end{center}
\end{figure}


The temperature dependence of the model parameters are shown in Fig.~\ref{fig8}(b). The amplitude of the modified Lorentzian $\chi_0$=$A_{ex}$ is proportional to the susceptibility of the spin fluctuations. This dynamic susceptibility gradually increases on cooling until it saturates and turns over at low temperatures. The remainder of the spin spectrum lies unresolved within the central mode discussed above. The data, however, shows that the dynamic AF correlations are still present within the bilayer at temperatures much higher than the temperature scale for appearance of the static central mode. The solid line in Fig.~\ref{fig8}(b) is the fit to a Curie-Weiss behaviour, $\chi_0= C/(T+\Theta_C)$, with $\Theta_C$ $\sim$32$\pm$8 K. As seen the data deviates from this behaviour below $\sim$30 K. The phenomenological Curie-Weiss temperature is considered the temperature scale for the appearance of short-ranged AF correlations at high temperatures in the paramagnetic state of geometrically frustrated magnets~\cite{gardner99} and some heavy fermions~\cite{lee01}.  

Fig.~\ref{fig8}(b) shows the relaxation rate decreases almost linearly on cooling down to about $\sim$80--100 K (dashed line) and eventually saturates to a constant value below $\sim$30 K. A linear temperature dependence for the relaxation rate is common in geometrically frustrated systems~\cite{gardner99} and it is also observed for lightly doped La$_2$Cu$_{0.94}$Li$_{0.06}$O$_4$~\cite{bao02} and La$_{1.95}$Ba$_{0.05}$CuO$_4$~\cite{hayden91}. In YBCO6.33 we find instead that a quadratic temperature dependence describes the relaxation rate, $\Gamma$(T)=a + bT$^2$, over the entire temperature range as shown with the dash-dotted line in Fig.~\ref{fig8}(b). 

Our analysis indicates that the inelastic spectrum cannot be described by a linear spin wave spectrum convoluted with the 4D resolution. In particular the inelastic peak cannot arise from the known vertical resolution. As shown later in Fig.~\ref{fig11} and discussed in relation to Fig.~\ref{fig13}, the finite extent in momentum of the spin correlations prevents divergence of the low energy susceptibility. Right at the staggered wavevector this results in a spin suppression leading to an overdamped peak centered on $\sim$3 meV.


A direct comparison of the temperature dependence of the intensity of low energy fluctuations with that of quasielastic scattering is shown in Fig.~\ref{fig9} where the intensity of the central mode and the excitations at 0.5 meV are plotted vs. temperature. Both data sets are collected with cold neutrons and are corrected for background. The central mode peak height grows on cooling while the low-energy fluctuations diminish. This shows that the excitations are indeed thermal fluctuations of the quasistatic ground state at low temperatures. This correlated behaviour also suggests that they arise from a single spin phase linked together by the hole doping, yet independent of the charge pairing.

Fig.~\ref{fig10} shows the integrated intensities of the inelastic and elastic scattering from thermal neutron data as a function of temperature. The inelastic integration is obtained from the constant-Q energy scans by numerical calculation of the area underneath the observed scattering from 1 meV up to 16 meV. This is justified since the Q-scans at different energies in this low energy range exhibit similar widths (see Fig.~\ref{fig11}). The elastic integration is obtained from elastic Q-scans in both H- and L-directions simply by multiplying the observed amplitude by the observed widths in both in-plane directions [HH0] and [-HH0], and by the width in the [00L] direction. The inelastic scattering decreases on cooling mirroring the increase in the elastic scattering. In this figure we have also shown the sum of the integrated intensities of elastic and inelastic components. The sum is temperature independent below $\sim$100 K as expected from conservation of the total moment. The decrease at higher temperatures arises because the scattering extends to much higher energies than our upper limit of integration of 16 meV.

\begin{figure}
\begin{center}
\vskip 0cm
\resizebox{.95\linewidth}{!}{\includegraphics{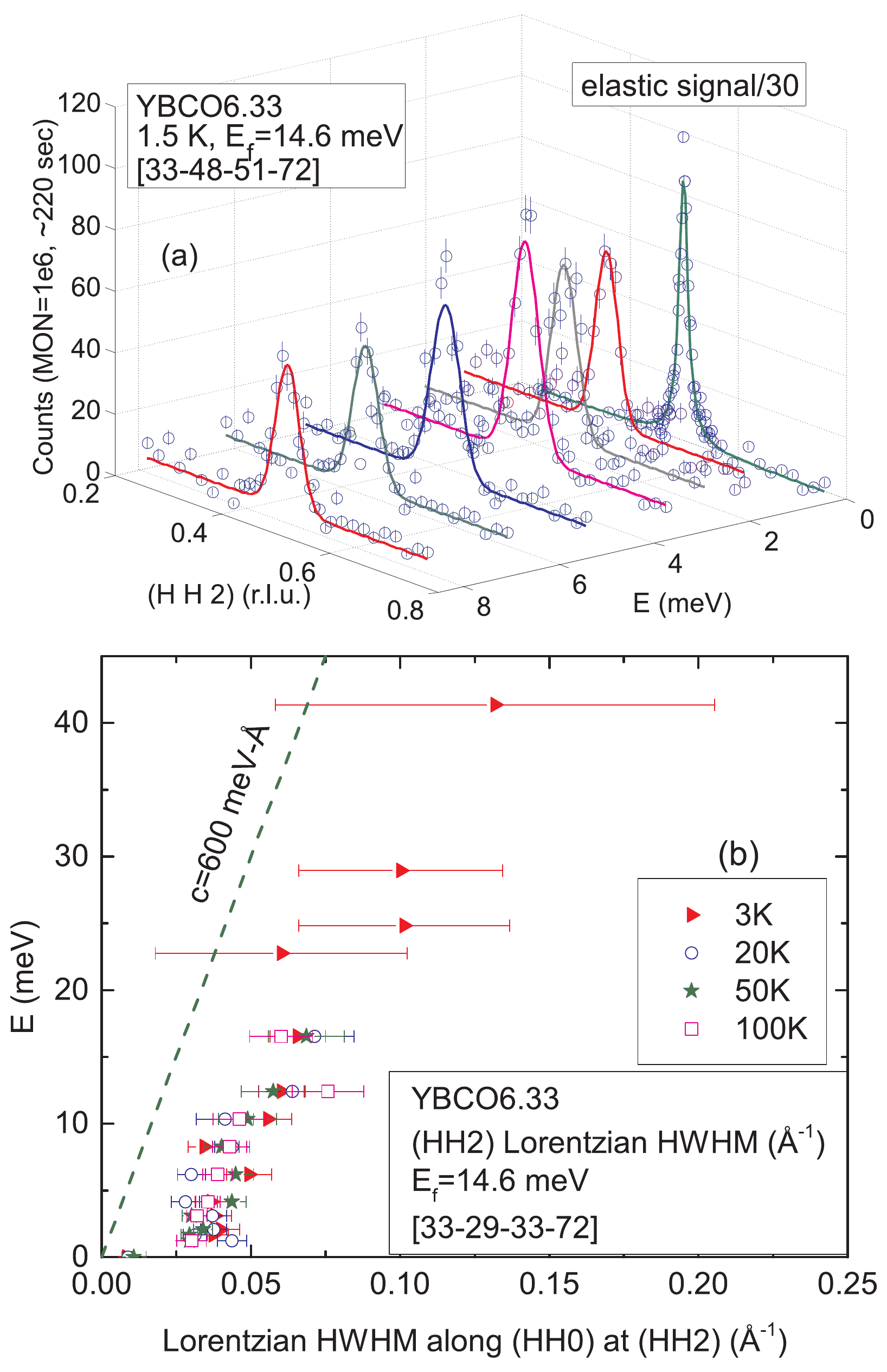}}\vskip
0cm \caption []{(a) The H-dependence of the scattering at different energy transfers around (0.5 0.5 2) at 1.5 K measured with thermal neutrons. (b) The resolution corrected HWHM in H of the inelastic scattering is constant and independent of temperature and energy from 1 to 15 meV and three times the elastic width (0.01 \AA$^{-1}$). The slope of the inverse spatial correlation length at high energies is similar to the spin-wave velocity~\cite{shamoto93} (broken line) of insulating YBCO6.15.   } \label{fig11}
\end{center}
\end{figure}

\begin{figure}
\begin{center}
\vskip 0cm
\resizebox{.95\linewidth}{!}{\includegraphics{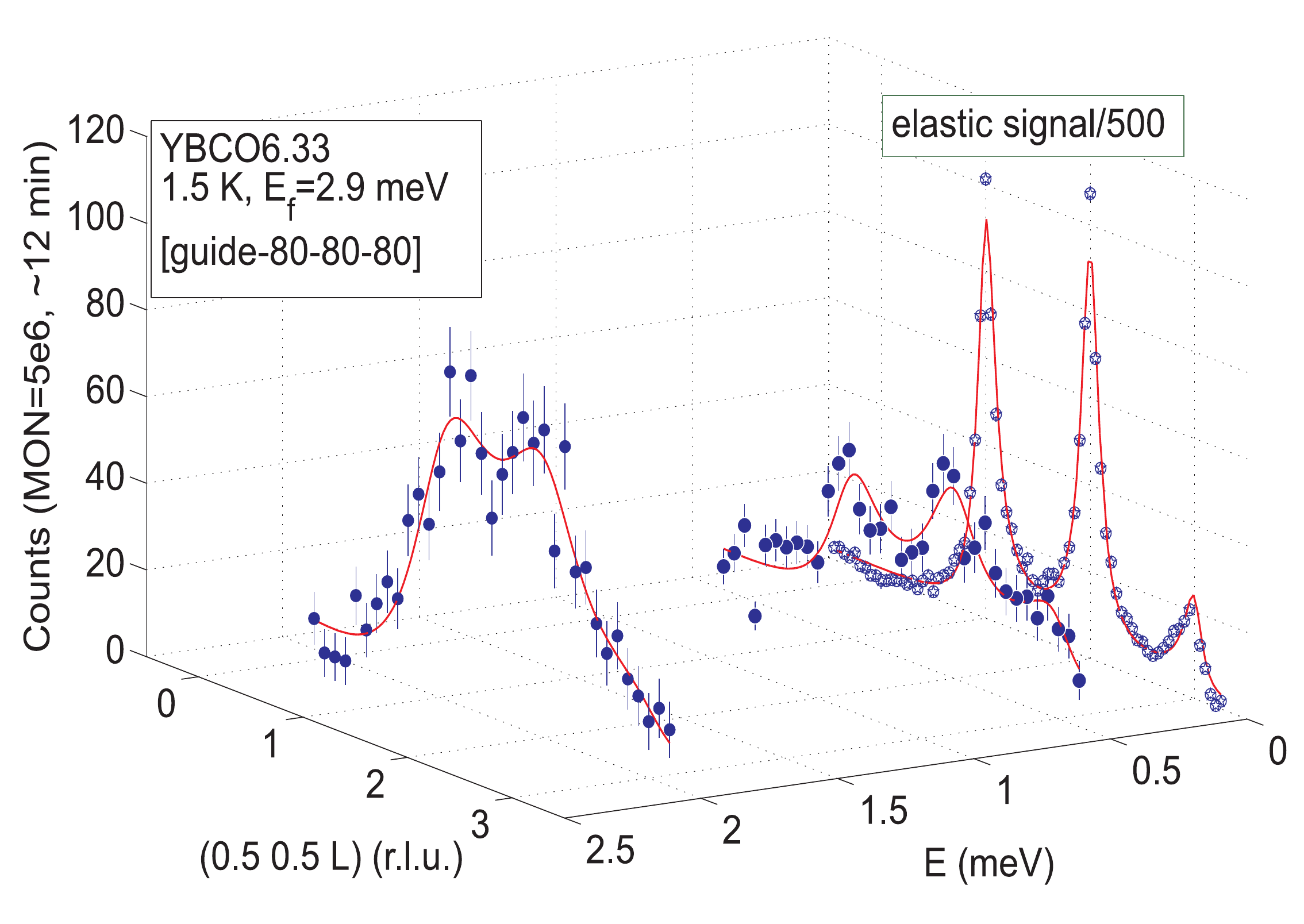}}\vskip
0cm \caption []{  The L-dependence of the scattering at 1.5 K at different energy transfers measured with cold neutrons.  The coupled bilayer modulation declines with energy and the L-dependence of the scattering becomes close to the independent bilayer form for E$\geq$2 meV. }\label{fig12}
\end{center}
\end{figure}
\begin{figure}
	\begin{center}
		\vskip 0cm
		\resizebox{.95\linewidth}{!}{\includegraphics{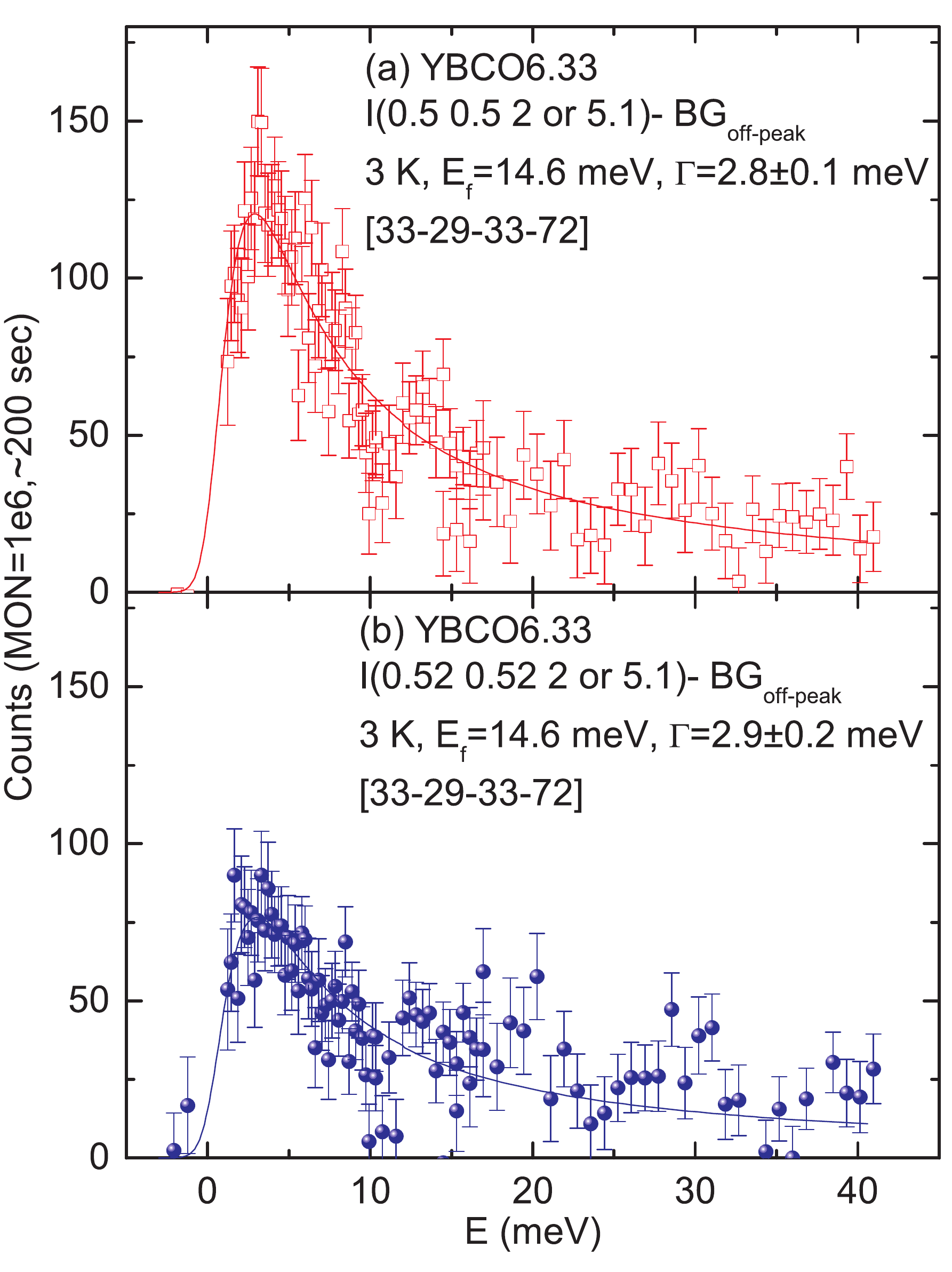}}\vskip
		0cm \caption []{Inelastic spectrum measured at (a) AF center $\mathbf{Q}_{\mathrm{AF}}$=(0.5 0.5 2) and (b) off the AF peak position (0.52 0.52 2) by twice the inverse correlation length (see Fig.~\ref{fig7}). In both panels high energy data is collected at (0.5 0.5 5.1) and corrected for the form factor. Again the average intensity observed at (0.3 0.3 2 or 5.1) and (0.7 0.7 2 or 5.1) is used as background, BG$_{\textrm{off-peak}}$. Modified Lorentzian with the same $\Gamma$ can describe both sets of data. Only the intensity is reduced at (0.52 0.52 2) compared with (0.5 0.5 2). } \label{fig13}
	\end{center}
\end{figure}

We have also performed constant-energy Q-scans along [HH0] in the energy range from zero to $\sim$40 meV and along [00L] from zero to 4 meV at different temperatures as shown in Figs.~\ref{fig11}(a), \ref{fig12} and \ref{fig5}(b). The half-width-at-half-max (HWHM) in {\AA}$^{-1}$, obtained from fitting the measured Q-scans along [HH0] to the convolution of resolution with the Lorentzian form of Eq.~\ref{QdepLor}, is shown in Fig.~\ref{fig11}(b) as a function of energy. For any nonzero energy where the central mode does not contribute, the fluctuations are three times as broad in H as the static response. Their widths remain constant at $\sim$0.03 {\AA}$^{-1}$ up to 15 meV above which they increase and tend to follow the spin-wave velocity measured~\cite{shamoto93} for the AF long-range ordered insulator YBCO6.15, c$_0$=600 meV\AA.

Constant energy Q-scans along [0.5 0.5 L] shown in Figs.~\ref{fig5}(b) and \ref{fig12} indicate that the integer-L centered coupled bilayer behaviour observed at zero energy is greatly reduced as energy increases. By 2 meV the dynamic correlations have lost the three-dimensional character leaving only the pattern for independent bilayers centered on L=1.7.

In order to investigate whether the relaxation rate rises with \textbf{Q}-\textbf{Q}$_{\mathrm{AF}}$, we measured inelastic scattering for nearby wavevector (0.52 0.52 2). Again we took the averaged intensity at (0.3 0.3 2 or 5.1) and (0.7 0.7 2 or 5.1) as background, BG$_{\textrm{off-peak}}$. Fig.~\ref{fig13} shows that the weaker off-peak fluctuations display the same relaxation rate independent of in-plane momentum H. Thus there is no dispersion associated with the low energy fluctuations in YBCO6.33. However the peak intensity at H=0.52 has fallen by only 1.3, much less that than the factor $\sim$6 expected for a Lorentzian dependence in q compared with the chosen offset of dH=0.02 r.l.u equivalent to dq =0.045 {\AA}$^{-1}$  where dq$\times$$\xi_{ab}$  = 4.5. The fact that the dynamic susceptibility holds up as q is displaced from \textbf{Q}$_{\mathrm{AF}}$ shows the tendency, even at this very low doping, for the spin strength to be slightly weighted towards incommensurate wavevectors as also suggested by Fig.~\ref{fig11}(b). Moreover, for all energy scans up to 40 meV there is no sign of any other spectral feature that can be associated with the so-called resonance observed in samples with higher doping. 


\begin{figure}
\begin{center}
\vskip 0cm
\resizebox{.85\linewidth}{!}{\includegraphics{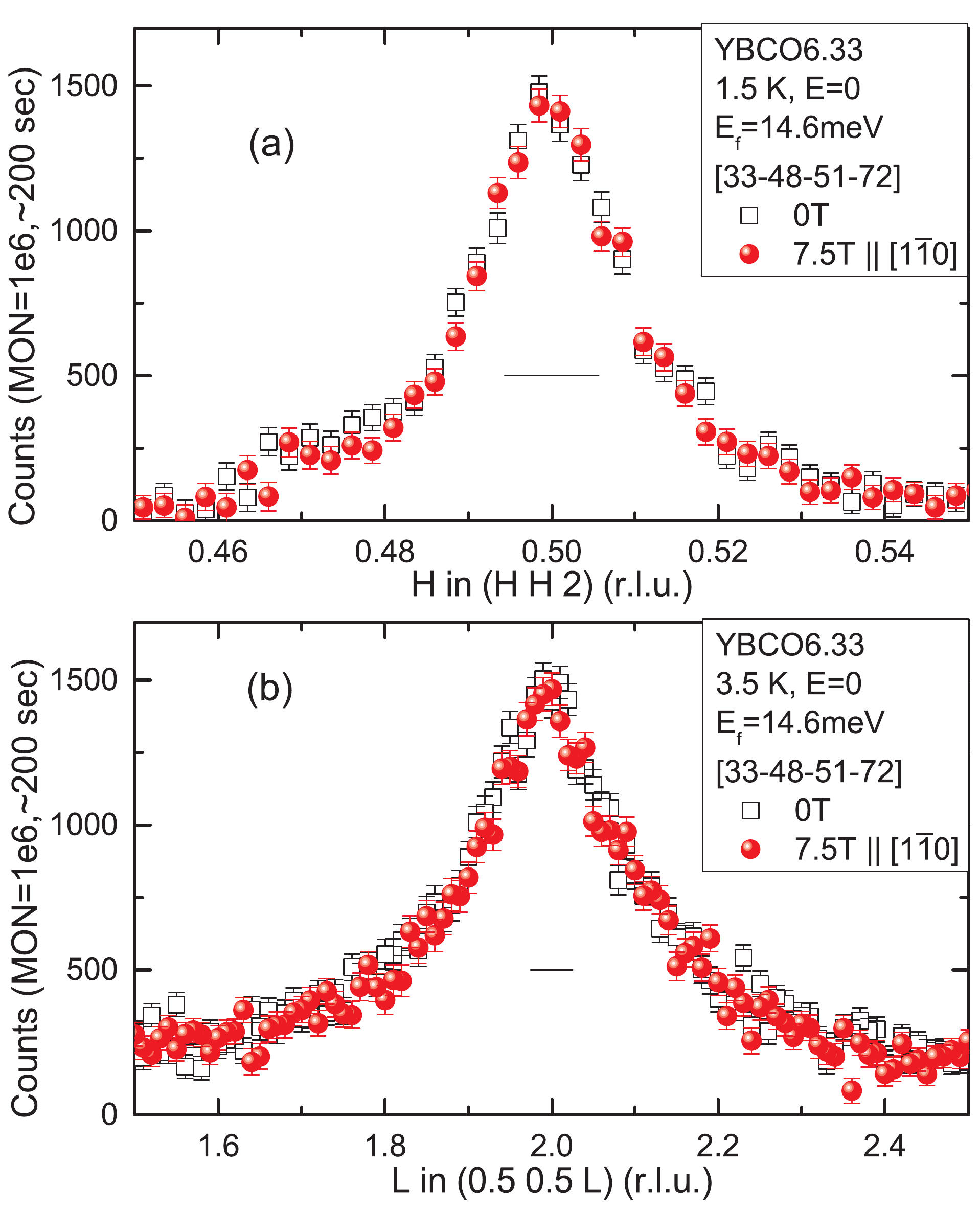}}\vskip
0cm \caption []{The elastic scattering does not change in an applied
field of 7.5 T parallel to [1$\overline{1}$0] as seen in scans (a) along [HH0] and (b) along [00L] through (0.5 0.5 2). The horizontal lines are the calculated resolution Bragg widths (FWHM).} \label{fig14}
\end{center}
\end{figure}

\begin{figure*}[tbh]
	\begin{center}
		\vskip 0cm
		\resizebox{.9\linewidth}{!}{\includegraphics{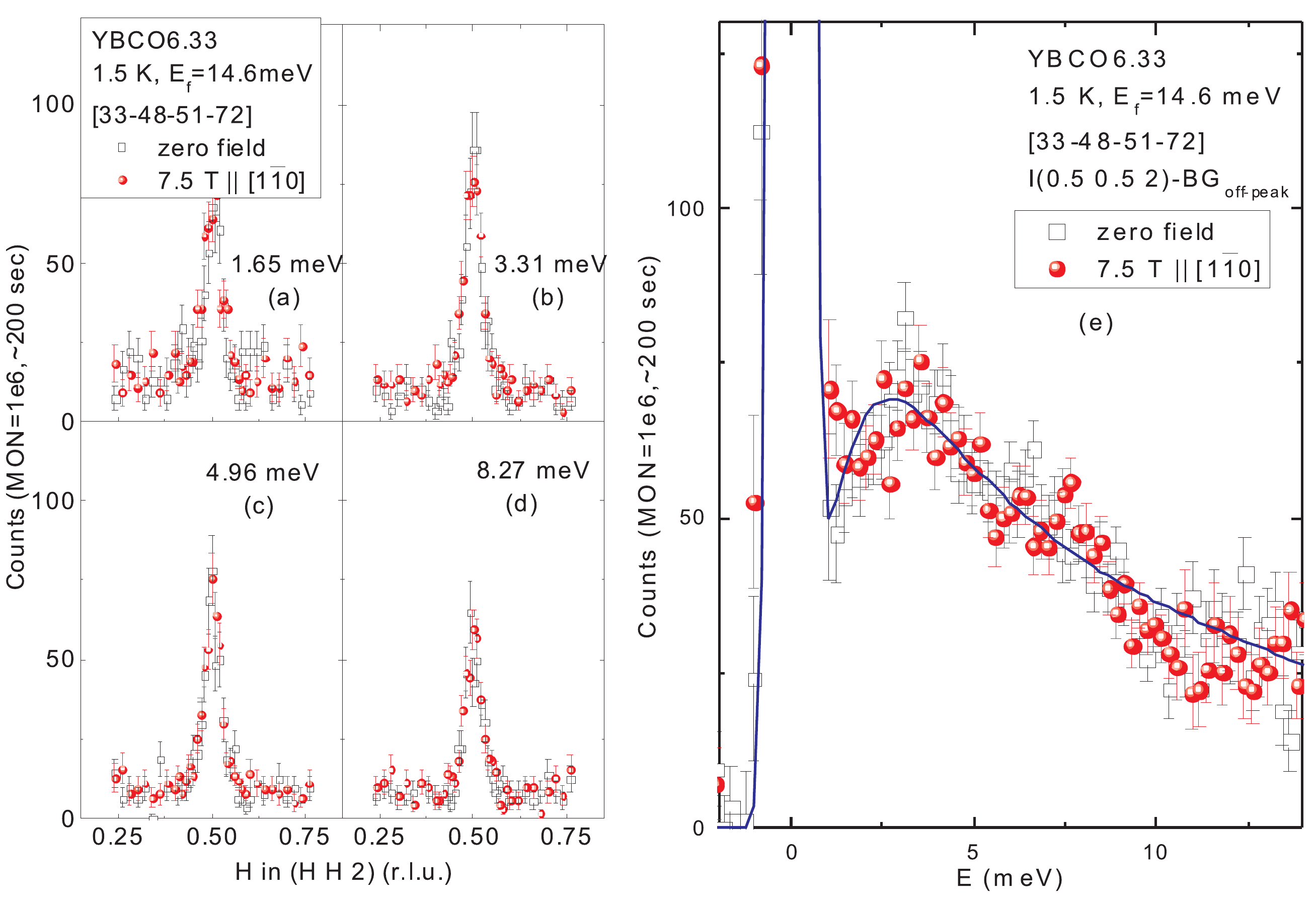}}\vskip
		0cm \caption []{(a)-(d) Constant-energy Q-scans at 1.5 K in an applied field of 7.5 T parallel to [1$\overline{1}$0] direction compared to zero-field. (e) The inelastic spectrum at 1.5 K measured at (0.5 0.5 2) in an applied field of 7.5 T field parallel to [1$\overline{1}$0] compared to zero field. Background again was taken as the average intensity at (0.3 0.3 2) and (0.7 0.7 2), BG$_{\textrm{off-peak}}$. No field dependence is observed for inelastic scattering similar to elastic scattering shown in Fig.~\ref{fig14}.} \label{fig15}
	\end{center}
\end{figure*}


\subsection{Magnetic field dependence}\label{magFdep}

Since one could argue that the transition to the superconducting state prevents further growth of the central mode at low temperatures (i.e. prevents a transition to a long-range AF state), we suppressed the superconductivity by applying a magnetic field of 7.5 T parallel to the [1$\overline{1}$0] direction. T$_c$ is reduced to 2.5 K from 8.4 K at 7.5 T as shown in Fig.~\ref{fig2}(a). Detailed measurements of the elastic scattering along [HH0] and [00L] directions through (0.5 0.5 2) at 1.5 K are shown in Fig.~\ref{fig14} and the temperature dependence of the central mode intensity is shown in Fig.~\ref{fig2}(b). Our standard method of high-temperature background subtraction was used. These measurements demonstrate that the central mode (intensity, q-profile and peak widths) and its temperature dependence are unaffected by the application of 7.5 T magnetic field, even though the system remains in its normal phase down to 2.5 K. It appears that the static spins in the central mode evolve separately from the charges that pair below 8.4 K.

In Fig.~\ref{fig15}(a-e) we show that the spin fluctuations both in q and in E are also unchanged by the applied field of 7.5 T. These observations and the absence of an anomaly in both elastic and inelastic scattering at T$_c$ confirm that superconducting properties play an unobservable influence on the very strong spin glassy correlations and their excitations. We do not probe the region near 0.8 meV, where a field enhancement at 6 T was seen~\cite{stock09} in YBCO6.35. This region is close to the strong elastic line and is hard to probe with the lower-resolution thermal neutrons. There is possibly very tentative evidence of a field enhancement near 1 meV (Fig.~\ref{fig15}e).


The lack of field dependence in both elastic and inelastic magnetic scattering in YBCO6.33 contrasts with the LSCO system~\cite{chang09,chang08,lake05,lake02,lake01,gilardi04,tranq04,khayko05}. In LSCO, for underdoped samples with a Sr concentration of x$\lesssim$0.15, the incommensurate magnetic scattering is enhanced~\cite{chang08,lake02} by application of a magnetic field. In the low doped superconducting La$_{2-x}$Ba$_x$CuO$_4$ with x=0.095 it was found~\cite{dunsiger08} that the incommensurate magnetic scattering is unaffected by magnetic fields up to 7 T applied along the c-axis. The lack of a field effect in this system compared to the presence of strong field dependence in LSCO was used to suggest~\cite{dunsiger08} that the field effect is not a universal property of cuprate superconductors. 

Experiments on more highly-doped YBCO6.45 with T$_c$=35 K have revealed~\cite{haug09} a strong enhancement of static incommensurate magnetic order at low temperatures by a magnetic field of $\sim$15 T applied mainly perpendicular to the CuO$_2$ planes. It was also found that the field reduces the amplitude of the inelastic response around $\sim$3 meV compensating for the spectral weight accumulated in the elastic peak. A similar field-induced suppression of the intensity is well known~\cite{dai00} for the resonance mode in more highly doped YBCO6.6. Based on these results it has been suggested~\cite{haug09} that the field enhanced magnetic superstructure is expected to drive a reconstruction of the Fermi surface which ultimately could explain the unusual Fermi surface topology revealed by recent quantum-oscillation experiments~\cite{dorion07}. Recent experiments on lower doped YBCO6.35 have instead shown~\cite{stock09} a magnetic field induced enhancement of the low energy spin fluctuations. This enhancement was suggested~\cite{stock09} to be the result of free spins located close to the hole-rich regions. Clearly further experiments are required in order to determine whether there is a common magnetic field dependence for the static and dynamic scattering as a function of doping in YBCO and other cuprates, but the initial results suggest otherwise.

\section{Discussion and Conclusions}\label{discussion} 

We find that in YBCO6.33 only short-range AF correlations are present. Despite the near-critical doping, the system lies inside the superconducting phase but outside any antiferromagnetic ordered phase. The spins fluctuate on two energy scales: one a damped spin response with a $\sim$3 meV relaxation rate and the other a resolution limited and intense peak at zero energy transfer (central mode). Even though we observe no transition to a long-range AF state, we demonstrate that the spins are highly correlated and exhibit an imprint of three-dimensional ordering with finite correlation lengths. 

The central mode develops gradually on cooling but does not diverge and the system remains subcritical. We find no evidence for a phase transition such as a N\'{e}el anomaly (Figs.~\ref{fig2}(b) and ~\ref{fig7}). The intensity of the central mode saturates on cooling below $\sim$10--15 K, with a concomitant minimum in the central mode peak widths. The presence of zero energy spin correlations over a finite wavevector fraction of the Brillouin zone is enough to prevent a transition to 3D long-range order. 

The central mode is centered on commensurate AF positions (0.5 0.5 L) with integral L indicating the development of weak 3D spin correlations along the c direction. The AF elastic correlations extend over only  $\sim$30 cells in the ab-plane and $\sim$1 cell along the c-axis. The short-range nature of the correlations can be understood by the AF frustration due to the doped holes. The doped holes produce extensive regions which break up the AF coupling and likely create ferromagnetic correlations between spins connected by oxygen neighbours in the CuO$_2$ planes where holes reside. This frustration leads to a spin-glass state. Such frustration becomes more significant at low temperatures when the ferromagnetic bonds become frozen and thereby exert a nonzero average field on the regions of AF correlated spins. The correlations are broadened on warming and the integral-in-L pattern becomes centered on the maxima of the bilayer structure factor, so indicating that 3D coupling between bilayers is vanishing. The avoidance of a quantum critical transition to 3D antiferromagnetism (divergence of the central mode intensity) is possibly related to breaking of the weakest coupling in the system, the antiparallel orientation of the spin of the upper member of the bilayer in one cell with the lower member of the bilayer in the next cell up along c. This is also evident from the shorter correlation length along c than the ab-plane.  

There have been several suggestions for the local texture that gives rise to the central mode. Haas et al. suggested~\cite{haas96} skyrmions but we find the skirt around the base of the central peak is hard to differentiate from our excellent Lorentzian fits. Some have suggested~\cite{ostland06,shraiman89,sushkov04,sushkov08,sushkov11} spin spirals. Aharony et al. suggested~\cite{aharony88} spin canting in the plane as frustration that destroys the N\'{e}el order at such a low doping. It is known that a doped hole, forms a Zhang-Rice singlet~\cite{zhangrice} that neutralizes a Cu spin. Since the doped hole extends over four or more oxygen sites, it is far more effective in destroying the long-range AF, reducing T$_N$ and magnetic moment, than simply a magnetic-site dilution. To date the local texture around a hole has not been established. 

\begin{figure}
	\begin{center}
		\vskip 0cm
		\resizebox{1\linewidth}{!}{\includegraphics{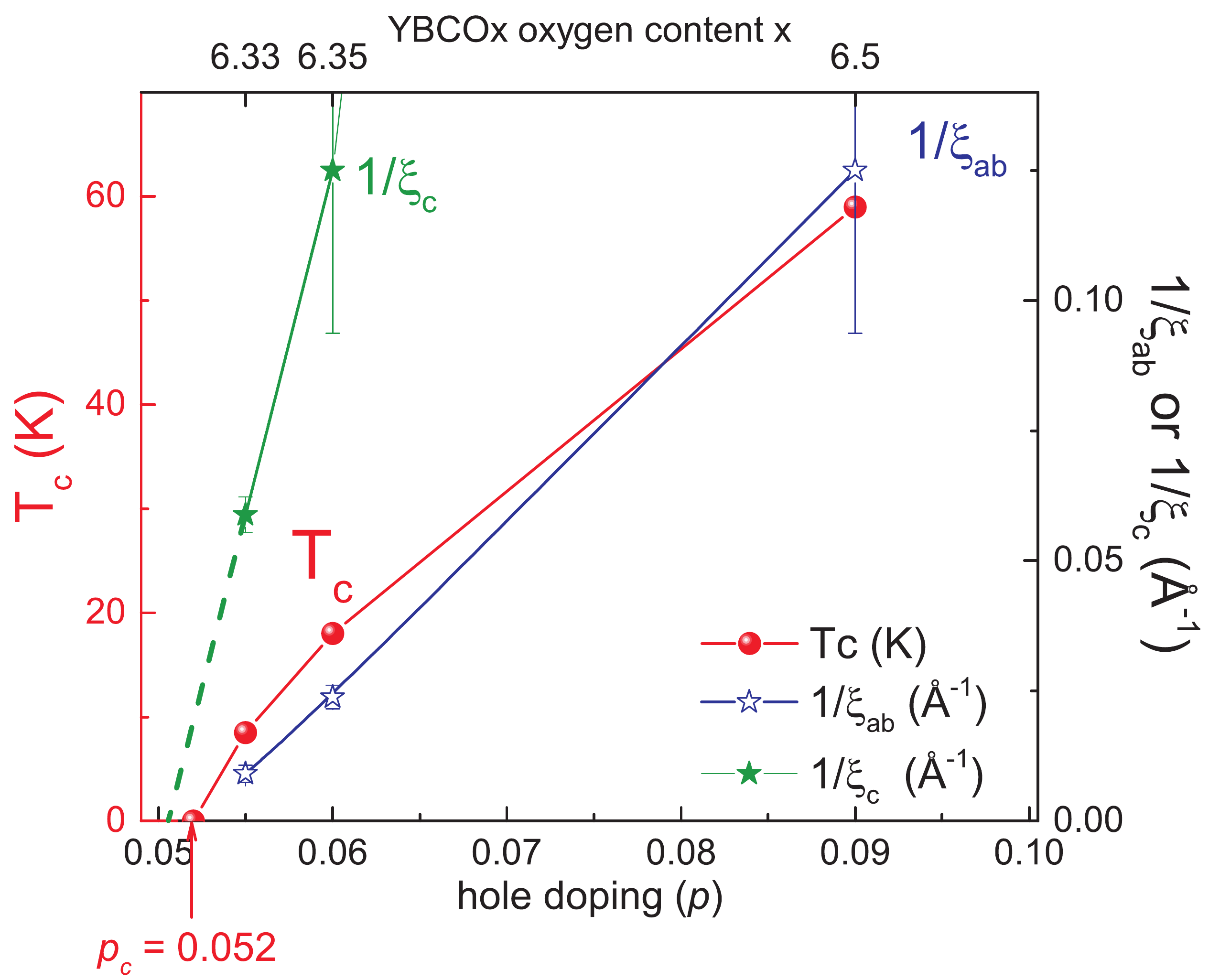}}\vskip
		0cm \caption []{The trend with doping of the correlation lengths in the ab-plane and along the c-axis are shown (right-hand y-axis). Doping dependence of superconducting transition temperature is also shown (left-hand y-axis). Data for YBCO6.35 and YBCO6.5 is taken from Ref.~\onlinecite{stock08}. As seen the correlation lengths become smaller as doping increases into the superconducting dome. For a small transition temperature of only 8.4 K in YBCO6.33, the correlation lengths remain finite.} \label{fig16}
	\end{center}
\end{figure}

The in-plane correlation length of about 100 \AA\ is much larger than the hole spacing expected for a 2D system with doping $p$=0.055, $l$=$a/\sqrt{p}$=16 \AA. The correlation seems to lie closer with $p$=0.055 to a critical behaviour $l$=$a/\sqrt{p-p_c}$=70 \AA. This is not unexpected as the spin system is becoming more three-dimensional as $p_c$ is approached. The trend with doping of the correlation range is shown in Fig.~\ref{fig16}. The inverse correlation lengths along both directions track T$_c$ without any sign of impending divergence at $p_c$. It is remarkable that reducing doping by only about 10$\%$ more than doubles the correlation lengths (see Table.~\ref{CorrvsDoping}) indicating high sensitivity of the static magnetic correlations to doping in this region. The closely coupled in-plane and out-of plane spin evolution we find suggests a highly organized structure rather than doping disorder. We also note that the development in q, $\omega$, and T of spin fluctuations is reminiscent of approach to a 3D phase transition and unlike what may occur in a system with disorder or random impurities. From a linear extrapolation of the $1/\xi_c$ observed for YBCO6.35 and YBCO6.33 (Fig.~\ref{fig16}), we suggest a true 3D long-range AF order to take place outside of SC dome at a doping lower than $p_c$=0.052. This suggestion is limited by the experimental accuracy, and is only possible because of the superb control of doping in the growth procedure used by the UBC group. Moreover there is little sign of a dynamic slowing of the spins as doping is reduced from YBCO6.35 to YBCO6.33. Evidently $p_c$ is a critical doping for the SC pairing, but not for the AF spin order.

\begin{table}
	\caption{Correlation lengths vs. doping for YBCO6.33 compared to YBCO6.35. Data for YBCO6.35 is taken from Ref.~\cite{buyers06}.} 
	\vskip .1cm
	\begin{tabular}{ c | c | c | c | c }
		\hline\hline
	~~Material~~               & ~~doping~~ & ~~T$_c$ (K)~~ &~~$\xi_{ab} (\AA)~~$~~ & ~~$\xi_{c} (\AA)$~~   \\
	\hline
	YBCO6.33 			& 0.055 & 8.4 & 110$\pm$20 \AA & 17$\pm$1 \AA \\
	YBCO6.35            & 0.06& 18 & 42$\pm$5 \AA & 8$\pm$2 \AA  \\
		\hline 
	\end{tabular} \label{CorrvsDoping} 
\end{table}

\begin{figure}[tbh]
	\begin{center}
		\vskip 0cm
		\resizebox{.9\linewidth}{!}{\includegraphics{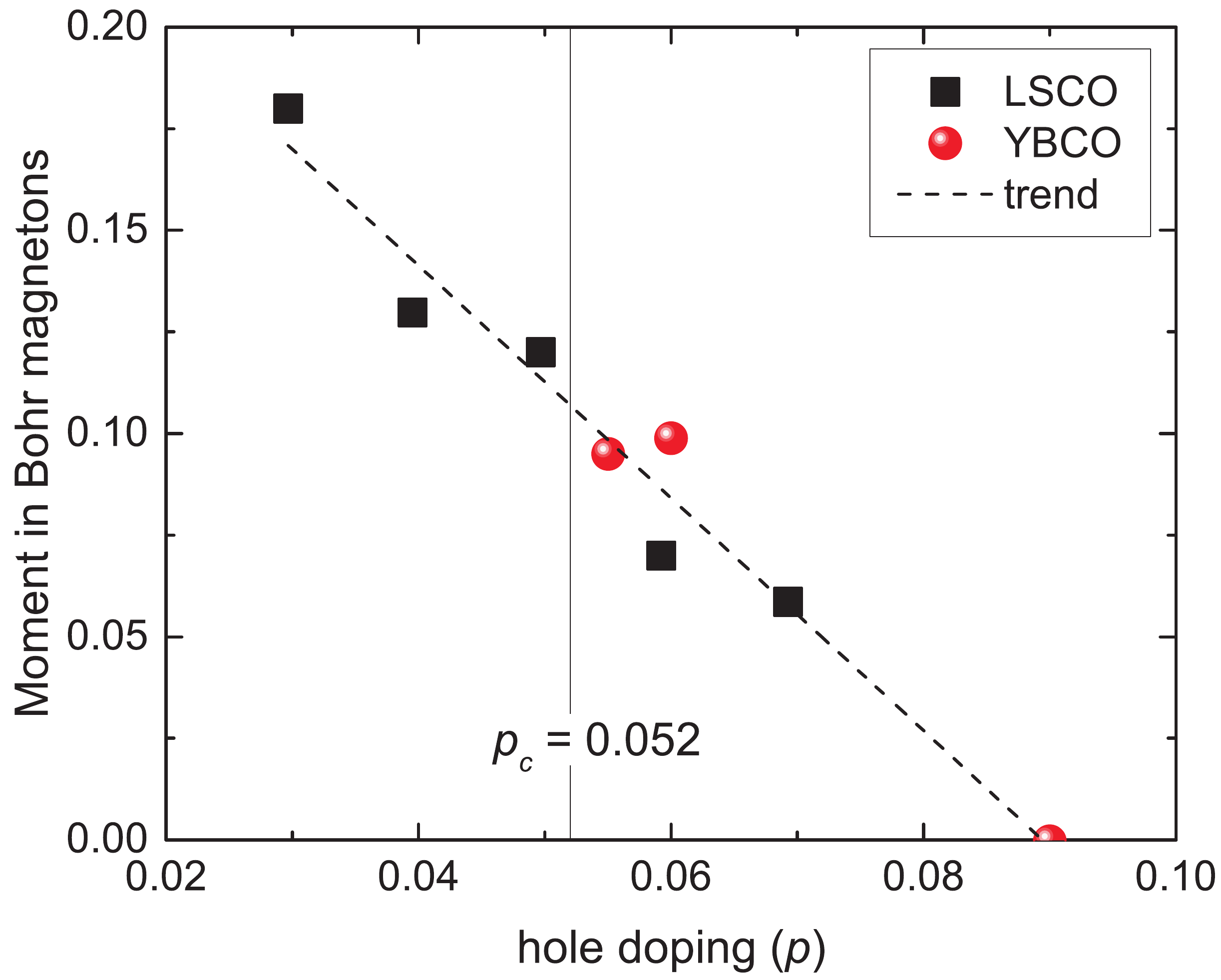}}\vskip
		0cm \caption []{The quasielastic moment, being the integral in q of the central mode, declines rapidly with doping. Data for LSCO system is taken from Ref.~\onlinecite{waki01}. Data for YBCO6+x with $p\geq0.06$ is taken from Ref.~\onlinecite{stock08}.  } \label{fig17}
	\end{center}
\end{figure}

The quasielastic moment, being the integral in q of the central mode at base temperature, declines rapidly with doping. Our results shown in Fig.~\ref{fig17} lie on a single trend line with doping that is the same for the YBCO and LSCO~\cite{waki01} systems. The value of the observed moment is derived by calibrating magnetic intensities from the observed intensities for several nuclear Bragg peaks~\cite{yamani10,stock08}. We have verified that integration of the elastic line when put on absolute scale by phonon calibration also results in a similar value to within error bars (0.08$\pm$0.02 $\mu_B$ compared to 0.11$\pm$0.02 $\mu_B$ from Bragg peaks). Their average is plotted in Fig.~\ref{fig17}.

Using high resolution probes (cold neutrons and the backscattering method) we have shown that while the central mode is never resolution limited in q (i.e. is not Bragg-like), it becomes resolution limited (less than $\sim$1 $\mu$eV) in energy domain at low temperatures. Thus the observed magnetic scattering appears static on a time scale of order of $\sim$10$^{-9}$ s. The central mode, being essentially elastic yet extending over a finite momentum range, breaks the relation of energy to momentum via velocity used to model highly-doped metallic cuprates~\cite{chou91}. The short-range correlations within and between planes reveal a ground state of frozen, sub-critical, 3D-enhanced spin correlations.   
 
We find the that the temperature scale for the growth of the correlations depends on the energy resolution of the probe (see Fig.~\ref{fig2}(b)). This behaviour observed for more conventional spin-glass systems~\cite{murani78,aeppli85,matsuda05} and also seen in the underdoped spin-glass phase of LSCO~\cite{matsuda00a} is taken as an indication that the observed scattering is not truly elastic. This is because a probe with a tighter energy resolution is more sensitive to slower fluctuations. It cannot detect the scattering at higher temperatures since the fluctuations are then faster than the time scale set by the energy resolution. The quasielastic characteristic of the scattering therefore can explain why the temperature scale of the central mode intensity depends on the energy window of the measurement. In YBCO6.33 magnetic correlations fluctuate with multi-relaxational rates and avoid a transition to a static order. From this behaviour and the fact that the correlation lengths remain finite, we deduce that the magnetic ground state of YBCO6.33 is spin-glass like. The spin-glass behaviour coexists with superconductivity and is not affected by the transition to the SC paired state below T$_c$ nor by magnetic fields in contrast to the LSCO~\cite{lake02,lake05} and LBCO~\cite{tranq14} systems. It appears that ordering of spins and SC paired charges proceed separately.
 
We do not observe a difference in the central mode intensity between field cooled and zero-field-cooled protocols as found in the spin-glass PbFe$_{1/2}$Nb$_{1/2}$O$_3$ by Chillal et al~\cite{chillal13}. This may be because of the much larger exchange in YBCO6.33 so that field effects are reduced below the resolution limit. Nonetheless our saturation of the peak intensity (below $\sim$10--15 K) does mirror the spin-glass behaviour observed by Chillal et al. Hence we view the spin-glass freezing temperature of YBCO6.33 as equal to $\sim$10--15 K since below this temperature, the intensity and the peak widths saturate to a low temperature value. The spin-glass freezing here is different from that in geometrically frustrated spin glasses but the concept can be applied to low doped YBCO6.33 since the spins are frozen in time and have no long-ranged order.
 
The absence of a transition to a long-range AF state differs from $\mu$SR results~\cite{sanna04,sanna10,niedermayer98} in underdoped YBCO6+x, Ca-doped YBCO and LSCO. The phase diagram constructed from such $\mu$SR measurements includes two magnetic transitions within the superconducting dome (a N\'{e}el transition at T$_N$ and a transition to a frozen spin glass at a lower temperature), where an ordered AF state is suggested to coexist with SC below T$_N$. Even though the sensitivity of $\mu$SR to slow spin fluctuations allows one to determine the onset temperature of magnetic correlations more easily and accurately, it only provides information on magnetic correlations integrated over q-space and therefore other complementary techniques such as neutron scattering are required to identify their spatial extent. We not only observe a gradual growth of the AF signal on cooling but also find that the correlation lengths in all directions remain finite to the lowest temperatures measured as is evident from a comparison of the observed magnetic peak widths with the experimental resolution and from a sample with long-range order (Fig.~\ref{fig4}). We note that the gradual onset temperature for appearance on the microvolt scale of the quasielastic central peak we observed with backscattering is comparable with T$_N$ determined from $\mu$SR measurements where the onset is sharp in T. 

We have determined that the magnetic correlations occur in form of broad but commensurate peaks. The LSCO family forms~\cite{kofu09,moho06,fujita02, matsuda02,waki01,matsuda00,matsuda00a,waki00,waki99,yamada98,keimer92} a spin-glass phase with short-range but incommensurate AF correlations coexisting with the SC phase for doping levels close to $p_c$. 
The incommensurability in the spin-glass phase scales linearly with doping. The lack of observable incommensurability for underdoped YBCO6.33 (this work) and YBCO6.35~\cite{stock08} is one of the differences that we find from the YBCO6+x (highly doped) and LSCO families. This may indicate that the presence of stripes is not crucial for superconductivity, although a breaking up of collinear spin order certainly is. It would have been hard to detect incommensurate correlations in a sample that is twinned and with disordered chains because the broadening would have obscured any anisotropy. Thus no attempt to do independent H and K scans was made to detect the cross-shaped scattering discussed by Gaw et al~\cite{gaw13}.

When comparing the properties of LSCO with YBCO, one must note the presence of intrinsic disorder caused by doping in LSCO~\cite{asini92}. The dopants occupy planes in close proximity to the active CuO$_2$ planes but not in YBCO. Perhaps the ideal system for investigating the intrinsic properties of low-doped superconductors with minimal disorder effects would be cuprates with a larger number of CuO$_2$ planes~\cite{tamaki12} since doping occurs well away from most of these planes. In fact recently it was reported that in a Hg-based superconductor with five CuO$_2$ layers where the effects of disorder are minimal, long-range antiferromagnetism and superconductivity coexist~\cite{mukuda08}. 

 \begin{figure*}[tbh]
 	\begin{center}
 		\vskip 0cm
 		\resizebox{.7\linewidth}{!}{\includegraphics{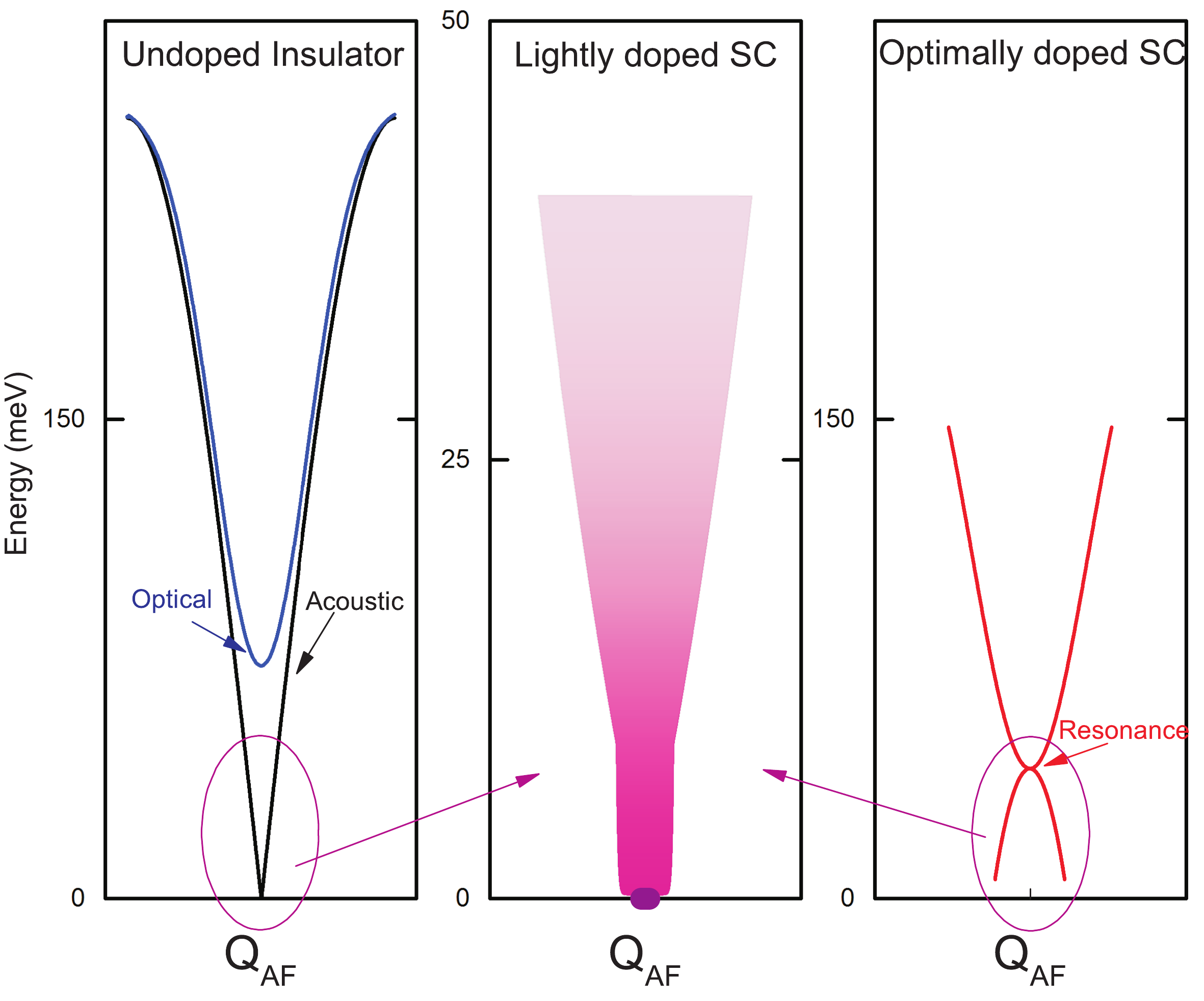}}\vskip
 		0cm \caption []{Magnetic excitations in undoped insulator (left), lightly doped SC YBCO6.33 (middle) and optimally doped YBCO (right) are schematically depicted.). Spin-wave spectrum observed for the insulator can be described~\cite{tranq07} by a Heisenberg Hamiltonian with strong anisotropic superexchange interactions.} \label{fig18}
 	\end{center}
 \end{figure*}

Our results are different from previous studies~\cite{keimer08, keimer10,keimer10a,haug09} on detwinned low-doped YBCO6.3 (T$_c$=0), YBCO6.35 (T$_c$=10 K) and YBCO6.45 (T$_c$=35 K) crystals where incommensurate elastic peaks with strong ab-plane anisotropy have been reported. These samples~\cite{keimer08, keimer10,keimer10a,haug09} were annealed for a short time ($\sim$1 day) and their SC transition temperatures are lower than the UBC crystals studied here which were annealed~\cite{stock05,stock04} for much longer ($\sim$3 months). Different planar doping for similar oxygen contents, therefore, may be the cause of observed differences. For a similar linear dependence of the static incommensurability vs. doping in LSCO along [1 0 0]~\cite{yamada98,fujita02}, the incommensurability for YBCO with a doping of 0.055 (our sample) is expected to be 0.055 r.l.u. in H along [1 1 0]. From elastic H-scans along [1 1 0] (Fig.~\ref{fig4}), we put an upper limit of 0.024 r.l.u. for possible static incommensurability in YBCO6.33, certainly lower than the expected value. For the inelastic peaks, we note that the FWHM of 0.058 r.l.u. observed in scans in H along [1 1 0] at 3 meV, if it arose from incommensurate peaks along pure [1 0 0] direction, would correspond to 0.058 r.l.u. separation of the putative incommensurate peaks. This is larger than the reported~\cite{keimer10} value of 0.025 r.l.u. for the dynamic incommensurability at 3 meV in YBCO with similar oxygen content. 

Even if our inelastic data may agree with the presence of dynamic incommensurability, from our elastic data, we can rule out the static stripe scenario for the very low doping of YBCO6.33 (T$_c$=8.4 K). This is in agreement with previous conclusions~\cite{stock06,stock08} for the slightly higher doped YBCO6.35 (T$_c$=18.5 K). It appears that for very low concentrations of hole doping in YBCO, an arrangement of antiferromagnetically correlated quasi-static spins in three dimensions over a finite range is more favored than alternating quasi-one-dimensional spin and charge regions. This may be due to the suppression of superconductivity compared to magnetism or the presence of only short-range oxygen chain order in very low doped YBCO. 

We find that the YBCO6.33 spin dynamics are relaxational with a relaxation rate, $\Gamma$, that saturates at the lowest temperatures. At high energies the dynamic planar correlations lengths shorten with an energy to q-width ratio similar to the spin-wave velocity in the insulating parent compound. Such a response in the energy domain (Eq.~\ref{modlor}) together with the Lorentzian momentum dependence (Eq.~\ref{QdepLor}) describes the local relaxation of antiferromagnetically coupled spins whose memory is lost after a characteristic correlation time ($1/\Gamma$) and over corresponding correlation lengths in the ab-plane and along the c direction. The finite relaxation rate as T$\rightarrow$0 reflects how the doped system avoids N\'{e}el condensation by having spin waves scatter off locally ferromagnetic regions surrounding doped holes. Therefore, this low energy suppression may be regarded as necessary for removal of spin fluctuations that might compete with superconducting order. There are thereby fewer pair-breaking spin fluctuations that match superconducting gap energies of order 3.5 k$_B$T$_c$=2.5 meV.  

The analysis of the temperature dependence of the magnetic excitations and the central mode indicates that $\sim$30 K can be regarded as the temperature scale for the appearance of the 3D spin-glass behaviour in YBCO6.33. Below this temperature not only does the strength of the damped fluctuations deviate from a Curie-Weiss law (Fig.~\ref{fig8}(b)) but also the central mode becomes more static and  three-dimensional (Fig.~\ref{fig7}(d,e,f)). This temperature is also close to the temperature below which quasistatic correlations gradually grow on cooling when measured with the very high energy resolution of backscattering (see Fig.~\ref{fig2}(b)).

Even though we observe the development of magnetic correlations at the AF zone centre on cooling, their damped spin response stands in stark contrast to the low-energy spin waves in the insulator and to the hour-glass dispersion of more highly doped superconductors (see Fig.~\ref{fig18} for a schematic comparison). While doping appears to have only a modest effect on the high energy excitations. The insulator undergoes a N\'{e}el transition to 3D long-range AF order where both the elastic scattering and the spin-waves can be described~\cite{tranq07} by an anisotropic exchange Heisenberg Hamiltonian. For our system, we observe instead a gradual increase of static short-range magnetic correlations with no low-energy spin waves. For higher doped YBCO6+x with x$\geq$0.45, the doped holes frustrate any AF order causing the entire scattering to be dynamic (no elastic magnetic scattering). The spectrum consists of a prominent resolution limited peak at the resonance energy, E$_{\mathrm{res}}$, located at the saddle point of the hourglass-shaped dispersion~\cite{li08,regnault95,fong00,dai01,hayden04,stock05,stock04,bourges97,bourges96,tranq07}. The observed hourglass magnetic spectrum and presence of incommensurate peaks around AF zone centre in cuprates are explained in terms of collective modes of an underlying stripe state~\cite{vojta06,andersen10,kivelson03,vojta09,norman01,rice06,uhrin04,vojta04,seibold05}.  

Since the only feature observed is the broad scattering at low energies ($\sim$3 meV), one might conclude that this overdamped Lorentzian is the low-energy remnant of a resonance in YBCO6.33. In fact from the empirical relation~\cite{bourges05} of E$_\mathrm{{res}}$=5.3k$_B$T$_c$ and a T$_c$ of 8.4 K, one might expect the resonance excitation to occur at 3.9 meV for YBCO6.33. Moreover, providing the integrated intensity of the resonance is related to the superfluid density~\cite{uemura06}, one might argue that because YBCO6.33 is such a weak superconductor (T$_c$ $\sim$8.4 K), the intensity of the resonance may be too small to observe. The modified Lorentzian we observe, however, does not display the properties typical of a resonance~\cite{regnault95,fong00,dai01,hayden04,stock04,stock05,li08} neither an increase in its intensity below T$_c$, nor an hour-glass dispersion nor a decrease with applied magnetic field (see Section~\ref{magFdep}). Instead the overdamped exciations decline on cooling as the spins drop into the static ground state revealed by the growing central peak. We, therefore, regard the overdamped scattering peaked at $\sim$3 meV as the spectral form of the excitations of an organized spin ground state with short-range correlations and incipient 3D response. Similar spin-glass behaviour is observed in lightly doped La$_2$Cu$_{0.94}$Li$_{0.06}$O$_4$~\cite{bao02}, La$_{1.95}$Ba$_{0.05}$CuO$_4$~\cite{hayden91} and La$_2$Cu$_{0.95}$Zn$_{0.05}$O$_4$~\cite{keimer92}.

A key point is that neither the growth in correlation length, the intensity of the central mode nor the spin dynamics show an anomaly at the superconducting transition temperature T$_c$=8.4 K, thus suggesting the spin correlations ignore superconductivity. This is fundamentally different from the higher doping (x$\gtrsim$0.4--0.5) region of the phase diagram. Since the transition to the superconducting state has no effect on the spin response, it appears that superconducting pairing of charges evolves independently of the glassy spin state as 3D magnetic correlations begin to dominate. This could be due to the fact that the pairs are already formed at temperatures much higher than T$_c$ or that the electrons responsible for superconductivity are different from those responsible for the magnetic scattering. The site-based N\'{e}el spin pairing evolves as the momentum-based spin pairing of charges is developing. 

The stark difference we see between the properties of the low-doped yet superconducting YBCO6.33 and the higher doped superconductors may be related to the presence of a critical hole concentration for a metal-to-insulator crossover (MIC) in the superconducting dome as suggested earlier~\cite{li08} from neutron scattering studies~\cite{li08} on YBCO6.45. The magnetic excitations were found to be gapless with a resonance much broader in energy than at the higher doping. A fundamental change in electronic properties within the SC dome can be similarly inferred from NMR~\cite{baek12} and transport measurements in a field~\cite{ando04}. The c-axis transport measurements~\cite{vignolle12} have also indicated that the coherence temperature for c-axis conductivity extrapolates to zero for doping close to $p\sim0.08$. Although charge order has now been observed~\cite{wu13} above a 10 T critical field in the vicinity of x=0.5 in YBCO6+x and for $p$=0.11 where the Ortho-II chain structure is present~\cite{blackburn13}, there is no X-ray report for charge order at the low oxygen contents such as in the YBCO6.33 we have studied. 

Although observations are qualitatively similar to higher doped YBCO6.35 (T$_c$=18 K)~\cite{stock06,stock08,buyers06}, these are the first measurements on YBCO6.33 with such a low T$_c$ revealing that even for such low doping, long-range antiferromagnetic order and superconductivity do not coexist. Rather the state coexisting with superconductivity is best regarded as a textured spin phase with short-range AF correlations. The measurements presented here on YBCO6.33 also enable us to predict that a transition to a true 3D AF state will most likely occur only outside the dome at a doping below $p_c$.  

Our results demonstrate that spin properties of very low-doped superconductors are fundamentally different from both the higher doped superconductors and the undoped insulators. Any microscopic theory for HTSC superconductivity must account for the strong doping dependence of the observed properties of spin dynamics.

\section*{Acknowledgements}

We are grateful to R. Sammon, C. Boyer, R. Donaberger, J. Fox, L. McEwan, D. Dean, S. Li,  T. Whan, M. Potter, and J. Bolduc at CNBC, Chalk River Laboratories, and the technical staff at NIST for excellent support. We acknowledge useful discussions with Zin Tun.

\onecolumngrid

\section*{Appendix: Absolute intensity scale}

\begin{figure}[tbh]
\begin{center}
\vskip 0cm
\resizebox{.45\linewidth}{!}{\includegraphics{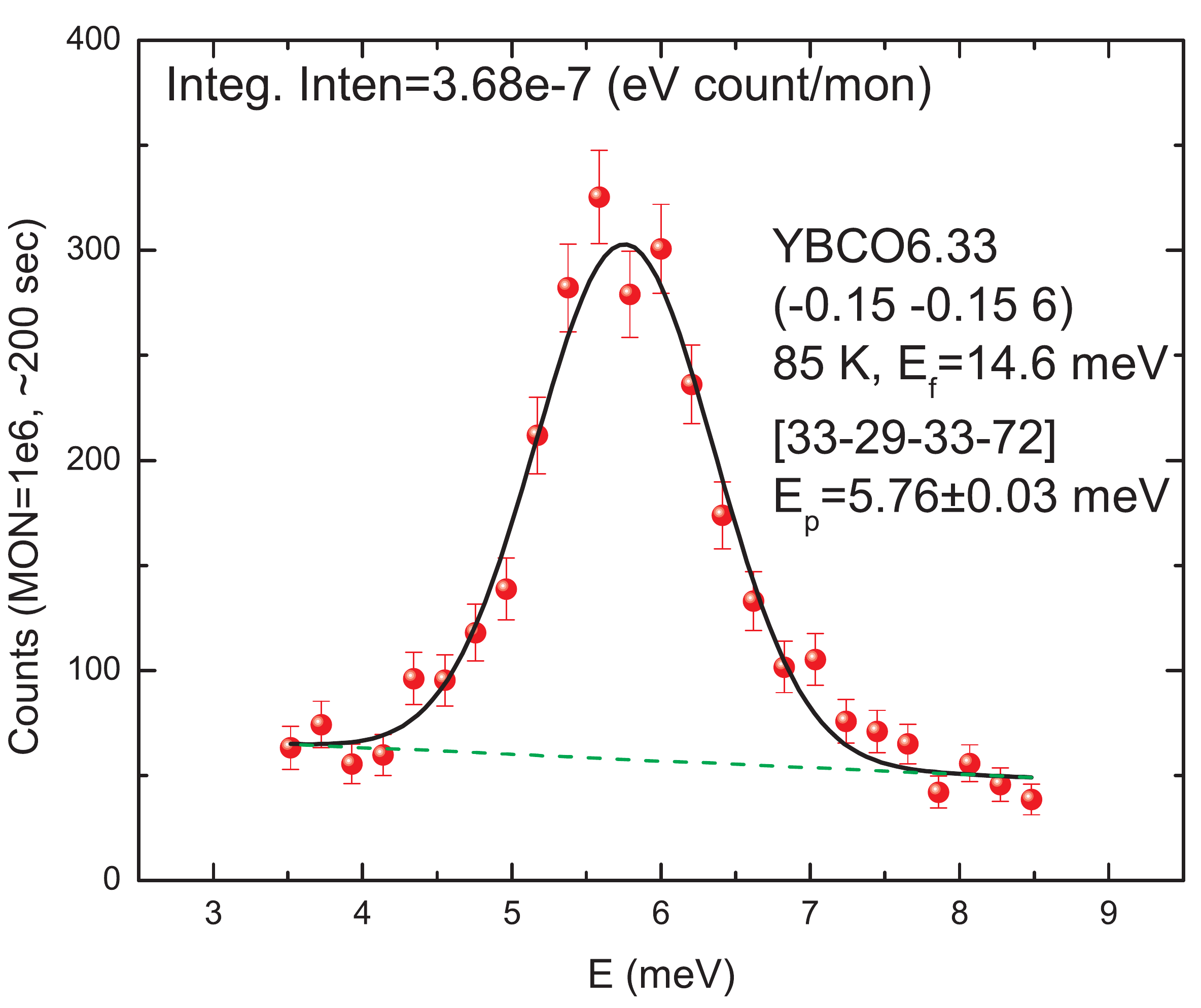}}\vskip
0cm \caption []{The transverse acoustic phonon measured at (-0.15 -0.15 6) close to nuclear Bragg peak (006) at 85 K. The fit is to a Gaussian function. The data shown is corrected for the presence of higher harmonics in monitor count rate. The integrated intensity of the peak above background is 3.68$\times$10$^{-7}$ (eV counts$/$mon).} \label{fig19}
\end{center}
\end{figure}

The measured magnetic scattering depends on the sample size and can be put on absolute scale using the observed integrated intensity of an acoustic phonon. This is because the cross section of acoustic phonon scattering in the long wavelength limit is known and can be easily calculated. In this experiment we measured the transverse acoustic phonon close to (006) at (-0.15 -0.15 6) at 85 K as shown in Fig.~\ref{fig19}.

The energy-integrated intensity, $I_{\mathrm{ph}}$(\textbf{Q})=$\int d\omega ~I(\mathbf{Q},\omega)$, for a coherent one-phonon creation process can be written~\cite{lovesey} as:

\begin{equation}\label{phononCS}
I_{\mathrm{ph}}(\mathbf{Q})= A~N~\left(\frac{\hbar^2 Q^2}{2M \hbar\omega_p}\right )~ |F_N(\mathbf{Q})|^2 ~e^{-2W\!(\mathbf{Q})}~ \mathrm{cos}^2\beta ~[1+n(\omega_p)]
    \end{equation}

\noindent where $A$ is the scale factor, $N$ is the number of nuclear unit cells, $\hbar\omega_p$ is the phonon energy, $[1+n(\omega_p)]=[1-e^{-\hbar \omega /k_B T]^{-1}}$ is the Bose factor, $e^{-2W\!(\mathbf{Q})}$ is the Debye-Waller factor (=1 in our calculations), $M$ is the mass of the unit cell, $|F_N(\mathbf{Q})|$ is the static nuclear structure amplitude of the Bragg reflection nearest to where the phonon is measured, and $\beta$ $\sim$0 is the angle between \textbf{Q} and the phonon eigenvector. From the data presented in Fig.~\ref{fig19} and Eq.~\ref{phononCS}, we find $A \times N$=2.04$\times$10$^{18}$(cm$^2/$eV-counts~per~mon). The constants used to obtain this ratio are given in Table~\ref{Phconstants}. 


The observed magnetic scattering intensity, $I_{\mathrm{mag}}(\mathbf{Q},\omega)$ is directly proportional to the magnetic scattering cross-section (see Eq.~\ref{magsection}) and can be written~\cite{lovesey} as:

\begin{equation}\label{magCS1}
I_{\mathrm{mag}}(\mathbf{Q},\omega)= A~N_{\mathrm{mag}} ~ g^2~(\frac{\gamma r_0 }{2})^2 |f(\mathbf{Q})|^2 ~B^2(Q_c)~e^{-2W\!(\mathbf{Q})}~  S(\mathbf{Q},\omega)
    \end{equation}
\noindent where $A$ is the scale factor, $N_{\mathrm{mag}}$ is the number of magnetic unit cells (=$N_n$ so that the absolute intensities are determined per  formula unit), $g\approx2$ is the Lande factor, $\gamma_n$=1.913 the gyromagnetic ratio of the neutron, $r_0$=2.817$\times10^{-13}$ cm the classical electron radius, $e^{-2W\!(\mathbf{Q})}$ is the Debye-Waller factor (=1 in our calculations), $|f(\mathbf{Q})|$ is the Cu$^{2+}$ anisotropic magnetic form factor~\cite{shamoto93}, and  $B(Q_c)$ is the bilayer structure factor in YBCO given~\cite{shamoto93} by

\begin{equation}\label{bilayerSF}
B(Q_c)=2~\mathrm{sin}( \frac{c Q_c }{2} z'_{\mathrm{Cu2}} )
    \end{equation}
\noindent where $Q_c=2\pi L/c$ and $z'_{\mathrm{Cu2}}=1-2z_{\mathrm{Cu2}}$ is the intra-bilayer spacing. The scattering function, $S(\mathbf{Q},\omega)$, is the Fourier transform of the spin-spin correlation function and is related to the spin susceptibility through the fluctuation-dissipation theorem

\begin{equation}\label{disfluc}
S(\mathbf{Q},\omega)= [1+n(\omega )] {\large \sum}_{\alpha,\,\beta} \left( \delta_{\alpha \beta}-\frac{Q_\alpha Q_\beta}{Q^2} \right ) \frac{\chi''_{\alpha \beta}(\mathbf{Q},\omega)}{\pi (g\mu_B)^2}
    \end{equation}
\noindent with the summation over Cartesian directions. Since there is no evidence for the presence of a long-range magnetic order and a preferred orientation of the moments, the summation in Eq.~\ref{disfluc} must be invariant with respect to the rotation of the indices. Therefore, the sum will be equal to ($2\chi''/ \pi (g\mu_B)^2$) where $\chi''$ is the isotropic susceptibility per  formula unit. Using this result, Eq.~\ref{magCS1} can be rewritten as

\begin{equation}\label{magCS2}
I_{\mathrm{mag}}(\mathbf{Q},\omega)= A~N_{\mathrm{mag}} ~ g^2~(\frac{\gamma r_0 }{2})^2 |f(\mathbf{Q})|^2 ~B^2(Q_c)~e^{-2W\!(\mathbf{Q})}~ [1+n(\omega )]~\frac{2\chi''(\mathbf{Q},\omega)}{\pi (g\mu_B)^2}
    \end{equation}

\noindent The constants used in the calculation of the absolute intensity at (0.5 0.5 2) are listed in Table~\ref{Magconstants}. The observed intensities of constant-Q energy scans, can now be put on an absolute scale using Table~\ref{Magconstants} and $A \times N$=2.04$\times$10$^{18}$(cm$^2/$eV-counts~per~mon) obtained above

\begin{equation}\label{magCS}
[1+n(\omega )]~\chi''(\mathbf{Q},\omega) =(3.38 \times10^{6})\times I_{\mathrm{mag}}(\mathbf{Q},\omega)
    \end{equation}

\noindent where $\chi''(\mathbf{Q},\omega)$ is in $(\mu_B^2/\mathrm{eV})$ per unit-formula and $I_{\mathrm{mag}}$ is in counts/mon. The absolute scales in Fig.~\ref{fig8}(a) are obtained from this method.


\begin{table}
	\caption{The constants used in the phonon absolute scale calculations. The transverse acoustic phonon data used in the calculation is shown in Fig.~\ref{fig16}. As discussed in the text $\mathrm{cos}^2\beta $=1. We have also approximated $e^{-2W\!(\mathbf{Q})}$ to 1. Lattice parameters used are a=3.844 \AA, b=3.870 \AA, and c=11.791 \AA.} 
	\begin{tabular}{c|c|c|c|c|c|c}
		\hline\hline
		\textbf{Q} &Q$^2$  & $|F_N(\mathbf{Q})|^2$ &$M$  &  $\hbar\omega_p$  &  $1+n(\omega_p)$ & $I_{\mathrm{ph}}^{(obs)}(\mathbf{Q})$   \\
		(r.l.u.)& (cm$^{-2}$) &  (cm$^{2}$) & (kg) &  (meV) & (unity) & (eV counts$/$mon) \\
		\hline
		(0 0 6) &1.02$\times$10$^{17}$& 1.73$\times$10$^{-23}$&1.09$\times$10$^{-24}$&  5.76& 1.84& 3.68$\times$10$^{-7}$\\ 
		\hline 
	\end{tabular} \label{Phconstants} 
\end{table}

\begin{table}[ht]
\caption{The constants used in the calculation of the absolute intensity of the magnetic scattering measured at (0.5 0.5 2). The number of magnetic unit cells is set equal to the nuclear unit cells $N_{\mathrm{mag}}$=$N$ so that the absolute intensities are determined per unit formula.} 
\begin{center}
\begin{tabular}{c|c|c|c|c}
	\hline\hline
  \textbf{Q} &$z'_{Cu2}$  & $(\frac{\gamma r_0 }{2})^2$  &  $|f(\mathbf{Q})|^2$  &  $B^2(Q_c) $   \\
  (r.l.u.)& (unity) & (cm$^2$) &   (unity) & (unity) \\
	\hline
 (0.5 0.5 2)& 0.36 & 7.27$\times$10$^{-26}$ &0.81& 3.86 \\
	\hline 
\end{tabular} \label{Magconstants} 
\end{center}
\end{table}

\twocolumngrid

\end{document}